\documentclass[aps,twocolumn,groupedaddress,showpacs,amssymb,prb,floatfix,preprintnumbers]{revtex4}
\usepackage{graphicx,color}
\usepackage{subfigure}
\usepackage{verbatim}

\begin{document}
\thispagestyle{myheadings}
\title{Magnetic properties and revisited exchange integrals of the
frustrated chain cuprate PbCuSO$_4$(OH)$_2$ - linarite}

\author{A.U.B.\ Wolter$^1$, F.\ Lipps$^1$, M.\ Sch\"{a}pers$^1$,
S.-L.\ Drechsler$^1$, S.\ Nishimoto$^1$, R.\ Vogel$^1$, V.\ Kataev$^1$,\\
B.\ B\"{u}chner$^1$, H.\ Rosner$^2$, M.\ Schmitt$^2$, M.\
Uhlarz$^3$, Y.\ Skourski$^3$, J.\ Wosnitza$^3$, S.\ S\"{u}llow$^4$,
K.C.\ Rule$^5$}

\address{$^1$Leibniz-Institut f\"{u}r Festk\"{o}rper und Werkstoffforschung IFW Dresden, P.O. Box 270116, D-01171 Dresden, Germany\\
$^2$ Max-Planck-Institut f\"ur Chemische Physik fester Stoffe, D-01171 Dresden, Germany\\
$^3$Hochfeld-Magnetlabor Dresden (HLD), Dresden-Rossendorf, D-01328 Dresden, Germany\\
$^4$Institut f\"{u}r Physik der Kondensierten Materie, TU
Braunschweig, D-38106 Braunschweig, Germany\\
$^5$Helmholtz-Zentrum Berlin f\"{u}r Materialien und Energie, D-14109 Berlin, Germany\\
}
\date{\today}

\begin{abstract}
We present a detailed study in the paramagnetic regime of the
frustrated $s$ = 1/2 spin-compound linarite, PbCuSO$_4$(OH)$_2$,
with competing ferromagnetic nearest-neighbor and antiferromagnetic
next-nearest-neighbor exchange interactions. Our data reveal highly
anisotropic values for the saturation field along the
crystallographic main directions, with $\sim$ 7.6, $\sim$ 10.5 and
$\sim$ 8.5\,T for the $a$, $b$, and $c$ axes, respectively. In the
paramagnetic regime, this behavior is explained mainly by the
anisotropy of the \textit{g}-factor but leaving room for an
easy-axis exchange anisotropy. Within the isotropic $J_1$-$J_2$ spin
model our experimental data are described by various theoretical
approaches yielding values for the exchange interactions $J_1$
$\sim$ -100\,K and $J_2$ $\sim$ 36\,K. These main intrachain
exchange integrals are significantly larger as compared to the
values derived in two previous studies in the literature and shift
the frustration ratio $\alpha = J_2/|J_1|$ $\approx$ 0.36 of
linarite closer to the 1D critical point at 0.25. Electron spin
resonance (ESR) and nuclear magnetic resonance (NMR) measurements
further prove that the static susceptibility is dominated by the
intrinsic spin susceptibility. The Knight shift as well as the
broadening of the linewidth in ESR and NMR at elevated temperatures
indicate a highly frustrated system with the onset of magnetic
correlations far above the magnetic ordering temperature
$T_\mathrm{N}$ = 2.75(5)\,K, in agreement with the calculated
exchange constants.
\end{abstract}

 \maketitle

\section{Introduction}

One-dimensional (1D) quantum magnetism involves the study of
materials with magnetic ions of low-spin state, $s$ = 1/2 or 1,
which are coupled magnetically along one crystallographic direction
only. Due to the rich nature of the low-temperature and
field-induced phases in these materials, quasi-1D quantum magnets
(Q1DQM) and their effective 1D models have attracted much interest
in the last decades from experimentalists and theorists
alike.~\cite{Schollwoeck2004,vyaselev04,clemancey06,laflorencie09}
Various Q1DQM exhibit highly fascinating field-induced phenomena
among which the Bose-Einstein condensation (BEC) of one-magnons
(triplons) \cite{jaime04,sebastian06,giamarchi08,kraemer07} or
two-magnon bound states near the saturation field
\cite{zhitomirsky10,svistov11} are particularly relevant examples.
Recent attention has been focussed on the latter phenomenon which is
well established for 1D models with competing interactions, where
via the inclusion of antiferromagnetic (AFM) next-nearest-neighbor
(NNN) interactions magnetic frustration plays a decisive role. A
prototype of these systems is the spin $s$ = 1/2 chain with
ferromagnetic (FM) nearest-neighbor (NN) and AFM-NNN interactions,

\begin{equation}
\hat{H} = J_1 \sum_{l} {\bf S}_l\cdot{\bf S}_{l+1} + J_2 \sum_{l}
{\bf S}_l\cdot{\bf S}_{l+2}-h \sum_{l} S_l^z, \label{Hamiltonian}
\end{equation}

\noindent with the NN exchange $J_1$ $<$ 0, the NNN interaction
$J_2$ $>$ 0, and $h$ as the external magnetic field along the $z$
direction. For these materials, it has been shown that the ground
state of the 1D Hamiltonian (1) has an instability towards
field-induced multipolar Tomonaga-Luttinger-liquid phases,
~\cite{sato09,sato10,kecke07,kuzian07,hikihara08,laeuchli09,dimitriev09,nishimoto10}
and which is interpreted as a hard-core Bose gas of multimagnon
bound states undergoing Bose-Einstein condensation at fields
slightly lower than the saturation field, $H_s$. The ground state of
the 1D Hamiltonian (1) is ferromagnetically ordered for $\alpha$ =
$|J_2/J_1|$ $<$ 0.25. At $|J_2/J_1|$ = 0.25, the FM state is
degenerated with a singlet state, while for $|J_2/J_1|$ $>$ 0.25 the
ground state is an incommensurate singlet. For a corresponding
quasi-1D system with small but finite interchain couplings, the
latter situation might possibly result in a helical magnetic
structure with an acute pitch (screw) angle. However, depending on
the strength of the exchange anisotropy and/or the spin-phonon
coupling, other types of helicoidal magnetically ordered states can
be realized, such as helical structures with obtuse pitch angles or
cycloidal helices, ordinary N\'{e}el states, a spin-Peierls phase,
or various other massive phases.~\cite{plekhanov08} Despite its
initially simplistic appearance, we would like to stress that the
$J_1$-$J_2$ model under consideration has not yet been fully
investigated theoretically, especially with respect to measurable
physical quantities. (To illustrate this point see e.g.\ Refs.\
[\onlinecite{kumar10, sirker11}] and references therein.)

The list of materials, which have been thought to be described well
by Eq. (1) and which could serve as a testing ground, includes
several quasi-one-dimensional (Q1D) edge-sharing chain cuprates,
such as Rb$_2$Cu$_2$Mo$_3$O$_{12}$, LiCuVO$_4$, NaCu$_2$O$_2$,
LiCu$_2$O$_2$, and
PbCuSO$_4$(OH)$_2$.~\cite{hase04,enderle05,buettgen07,buettgen10,svistov11,
drechsler06,masuda04,masuda05,park07,seki08,kamieniarz02,baran06,yasui11}
For most of these systems the magnetic couplings -- and thus $H_s$
-- are rather large provided the system is not in the vicinity of a
ferromagnetic critical point such as Li$_2$ZrCuO$_4$. In the latter
case, single crystals are still not available making it generally
difficult to study effects close to $H_s$ experimentally using
stationary fields.~\cite{drechsler07,drechsler07a,
schmitt09,vavilova09}

\begin{figure}
\begin{center}
\includegraphics[width=0.85\columnwidth]{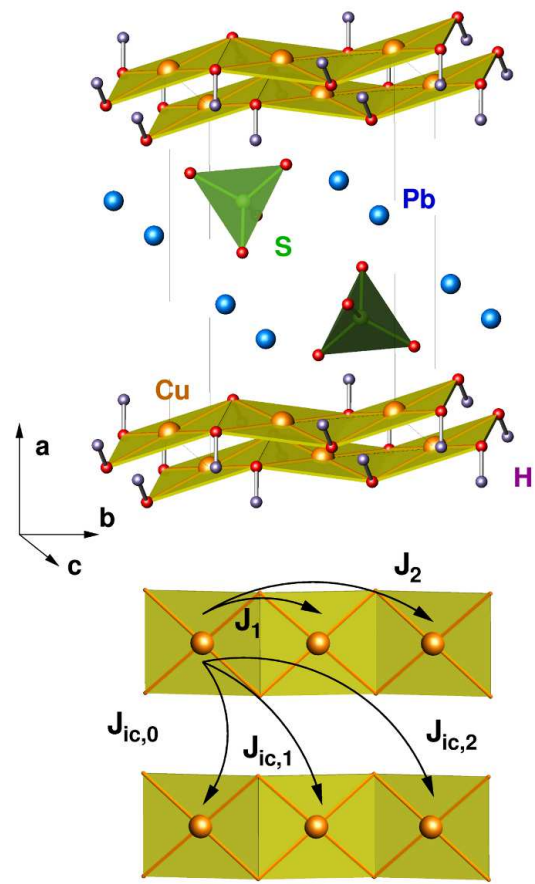}
\end{center}
\caption[1]{(Color online) The crystallographic structure and the
main exchange paths in PbCuSO$_4$(OH)$_2$. The CuO$_2$ units are
aligned in the bc plane, forming edge-sharing CuO$_2$ chains along
the $b$ direction. In order to illustrate the coordination of the S
atoms, oxygen tetrahedra are highlighted in the sketch. The two
inequivalent alternating "left" and "right" proton positions
distinguished by bond lengths and bond angles are according to
Ref.~\onlinecite{schofield09}.} \label{fig:structure}
\end{figure}

Possibly, only one of these materials, the natural mineral linarite,
PbCuSO$_4$(OH)$_2$, reflects an optimum compromise of a
comparatively low $\alpha$-value not far from the critical point,
{\it viz.}, a low saturation field, and the availability of single
crystals. Despite this, linarite has only been studied superficially
to date. According to these initial studies, it has been proposed to
represent a frustrated quasi-1D $J_1 - J_2$ model magnet described
by Eq.~(1). Linarite crystallizes in a monoclinic lattice (space
group \textit{P2$_1$/m}; $a = 9.682$\,\AA, $b = 5.646$\,\AA, $c =
4.683$\,\AA, $\beta$ = 102.66$^{\circ}$),~\cite{schofield09} in
which CuO$_2$ units are aligned in chain-like structures along the
$b$ direction (Fig.~\ref{fig:structure}). Each of the Cu$^{2+}$
atoms is fourfold coordinated, with the surrounding oxygens forming
a flat tetragon close to a square, the celebrated CuO$_4^{-6}$
plaquette-"brick" of all undoped cuprates. This is supplemented by
four hydrogen atoms as ligands to these planar oxygens. The Cu
coordination is completed by two further oxygen atoms from the
SO$_4$ groups, yielding a distorted octahedron and a slightly
non-planar "waving" (buckled) CuO$_2$-chain structure. This is in
contrast to the much better studied planar counterpart of the
edge-shared chain cuprates LiCu$_2$O$_2$ and LiVCuO$_4$.

Initial susceptibility and zero-field specific heat data have been
interpreted in terms of a dominant magnetic coupling along the
chain, with a {\it predominant} FM-NN interaction $J_1 = -30$\,K and
a weaker AFM-NNN coupling $J_2 = 15$\,K ($\alpha$ =
0.5).~\cite{baran06} This way, a strong coupling scenario in terms
of interacting and interpenetrating simple AFM Heisenberg chains
would be envisaged for the $J_1$-$J_2$ model under consideration.
However, more recent studies by Yasui et al.~\cite{yasui11} indicate
a rather different {\it weak}-coupling regime, namely: $J_1$ = (-13
$\pm$ 3\,)\,K and $J_2$ = (21 $\pm$ 5\,)\,K ($\alpha$ = 1.6), which
have been obtained from a fit to the susceptibility data using a
high-temperature expansion up to the fourth-order in the temperature
range 50 $<$ $T$ $<$ 350\,K. Such a weak-coupling regime should give
rise to more pronounced frustration and fluctuation effects.
\cite{nishimoto11} The observation of a magnetically ordered state
below $T_\mathrm{N}\approx 2.8$\,K has been discussed in terms of a
possible helical ground state with an acute pitch
angle.~\cite{baran06} The helical nature of the ground state is
supported by a recent study of the dielectric constant in this
material.~\cite{yasui11} However, microscopic studies such as NMR or
neutron scattering measurements to prove these predictions are still
lacking. Although in both references a similar conclusion has been
obtained regarding the basic nature of the magnetic ground state of
linarite,~\cite{baran06,yasui11} the physical characteristics of the
spiral and thus the magnetic properties would differ enormously due
to e.g. the relevance of quantum effects on the pitch angle and on
the magnetic moment in the ordered state.

In view of these conflicting results, we started a detailed study on
this material combining macroscopic and microscopic experimental
techniques with different theoretical methods to resolve the
magnetic exchange parameters of PbCuSO$_4$(OH)$_2$ and, thus, its
underlying ground state. Measurements of the static susceptibility
and the saturation magnetization have been performed. The intrinsic
spin susceptibility is investigated utilizing ESR and NMR.
Measurements of the $g$-factor as well as of the ESR and NMR
linewidths indicate an appreciable magnetic anisotropy in our system
and further emphasize the highly frustrated character of linarite
with an onset of magnetic correlations far above the magnetic
ordering temperature. An analysis of our data within advanced
theoretical methods, such as density matrix renormalization group
(DMRG), hard-core boson technique and local (spin-polarized) density
approximation (L(S)DA+$U$) calculations, which all take into account
the nontrivial interplay of quantum effects and frustration beyond
linear spin-wave theory, yield new values for the exchange couplings
along the chain directions of $J_1$ $\sim$ -100\,K and $J_2$ $\sim$
36\,K. These values are substantially larger than those determined
previously.~\cite{baran06,yasui11} Within our extensive analysis, we
succeeded in estimating the order of magnitude of the interchain
couplings. The theoretical part of the present paper should be
understood also as a step in the direction to fill the existing gap
of the not fully explored $J_1-J_2$ model, beyond providing an
assignment of the main exchange parameters for a specific material
only. In particular, a precise knowledge of the main exchange
parameters is one of the prerequisites to attack in a realistic
manner rather complex and not yet fully understood phenomena such as
multiferroicity \cite{yasui11} and multipolar phases/spin nematics
\cite{sato10} reported or suggested for the title compound, too.

\section{Methods}

\subsection{Experimental}

\subsubsection{Samples and diffractions}

The single crystals of PbCuSO$_4$(OH)$_2$ used in this study for the
magnetization, NMR, and X-band ESR measurements are natural minerals
with their origin in California, USA (\emph{origin 1}: Blue Bell
Mine, Baker, San Bernadino). A second set of naturally grown single
crystals of smaller size (\emph{origin 2}: Siegerland, Germany) has
been used for the ESR measurements in a resonant cavity at a
frequency of about 93\,GHz. All crystals show well-defined facets
and the principal axes $b$ and $c$ can be identified easily. Single
crystallinity of our samples has been checked by x-ray diffraction.
For both sets of single crystals no magnetic impurity phases were
observed within experimental resolution, as evidenced by the absence
of a low-temperature Curie tail in the magnetic susceptibility. For
all measurements the samples were oriented along the principle
crystallographic directions with a possible misalignment of less
than 5$^\circ$.

\subsubsection{Magnetization}

Temperature-dependent magnetization studies of linarite were
performed using a commercial SQUID magnetometer in the temperature
range 1.8 - 400\,K and in an external magnetic field of 0.4\,T.
Magnetization curves, $M(H)$, were measured at $T$ = 1.8 and 2.8\,K
in a commercial Physical Properties Measurement System using a
Vibrating Sample Magnetometer (VSM). Magnetization data were
collected while sweeping the magnetic field using sweep rates of
about 50\,Oe/s for both increasing and decreasing field regimes.
Note that due to hysteresis around the phase transitions observed
for the 1.8\,K data, the sweep rate was significantly varied in
these regions in order to check for sweep-rate dependent effects on
the phase transitions. Using quasi-static conditions, the observed
small hysteresis in the $M(H)$ curves became negligible, as shown
below.

\subsubsection{Electron Spin Resonance (ESR)}

For the ESR experiments two different setups were used. The
commercial X-band spectrometer operates at a frequency of about
9.6\,GHz. This allows sweeps of the magnetic field up to 1\,T. The
setup is equipped with a continuous-flow liquid-helium cryostat,
enabling measurements from room temperature down to about 4\,K. The
cryostat is inserted in a rectangular microwave resonator in the
TE102 mode configuration. Samples are mounted on a quartz sample
holder, which is centered in the resonator at the maximum of the
microwave magnetic field. The sample holder can be rotated with
respect to the external magnetic field. By using an additional
external magnetic modulation field, the lock-in detection technique
is applied. The second setup consists of a microwave vector network
analyzer (MVNA) and a 15 Tesla superconducting
magnet.~\cite{Golze06} This setup allows phase-sensitive
measurements at different frequencies in the range from 30-800\,GHz.
Most experiments were carried out using a home-built cylindrical
resonator in the TE011 configuration with a resonance frequency of
about 93\,GHz. For this, the samples were mounted on a quartz needle
in the center of the resonator. The resonator is coupled to a
metallic waveguide, which is placed in the center of the
superconducting magnet. The very high quality factor of the order of
$Q\sim10^{4}$ -- and, thus, the high sensitivity of this resonator
setup -- allowed the measurement of single crystals of linarite with
dimensions of approximately 0.5 x 0.5 x 1\,mm$^3$ with very high
signal-to-noise ratio.

\subsubsection{Nuclear Magnetic Resonance (NMR)}

$^1$H-NMR ($^1\gamma$ = 42.5749 MHz/T) and $^{207}$Pb-NMR
($^{207}\gamma$ = 8.9074 MHz/T) measurements were performed using a
phase-coherent Tecmag spectrometer with a He-flow cryostat for
temperatures down to 4.2\,K and magnetic fields of 2 and 4\,T,
respectively. Temperatures below 4.2\,K were achieved by pumping on
the helium bath. The NMR spectra were determined using a $\pi$/2 -
$\tau$ - $\pi$ Hahn spin-echo pulse sequence. Special care was taken
to avoid extrinsic signals from parasitic $^1$H atoms around the
sample. Due to the strong increase of the linewidth at low
temperatures, additional field sweeps at constant frequency have
been performed. This way, we can ensure that neither the selected
frequency excitation spectrum by the pulse width (typically a
$\pi$/2-pulse is about 3 $\mu$s) nor the quality factor of our coil
do artificially narrow the detected lines of PbCuSO$_4$(OH)$_2$. The
spin-lattice relaxation time $T_1$ has been recorded using an
inversion-recovery pulse sequence ($\pi$ - $\tau_{var}$ - $\pi$/2 -
$\tau$ - $\pi$) with variable delay $\tau_{var}$ and a Hahn
spin-echo detection at the end. Typical conditions of excitation
were 3\,$\mu$s and 6\,$\mu$s for a $\pi$/2- and $\pi$-pulse,
respectively. Repetition rates were in the range 100 - 400\,ms
despite short spin-lattice relaxation times $T_1$ in order to avoid
any local heating at the sample site. For $H$ $||$ $b$, the
spin-lattice relaxation rate of the $^1$H nucleus was determined
using a saturation-recovery sequence with an echo subsequence at the
end, i.e., ($\pi$/2 - $\tau_{del}$)$_n$ - $\tau_{var}$ - $\pi$/2 -
$\tau$ - $\pi$ with the delay time $\tau_{del}$ and $n$ as the
number of repetitions of the first cycle. Calibration of the fields
has been performed using the $^1$H- and $^2$D-NMR resonance
frequencies of hydrogenated and deuterated water at room temperature
for the 2 and 4\,T experiments, respectively.

\subsection{Theoretical methods}

\subsubsection{DMRG and TMRG}

Initially, we analyzed the saturation field using well-known
rigorous expressions valid in the so-called two-magnon and
one-magnon sectors depending on the strength of the interchain
coupling. Next, we considered the magnetization curve at low $T$,
say 1.8\,K, for $H$ $\perp$ $(bc)$ and compared our calculations
with the experimental data shown in Fig.~\ref{fig:magnetization} for
this direction. To calculate the magnetic susceptibility we employed
the transfer-matrix renormalization group (TMRG)
method.~\cite{Bursill1996,Wang1997,Shibata1997} In our calculations,
80-160 states were retained in the renormalization procedure and the
truncation error was less than 10$^{-4}$ down to $T$ = 0.003$|J_1|$.

To calculate the static spin-structure factor
$S(q)=(1/L^2)\sum_{ij=1} ^L \left[ \langle S_i ^zS_j ^z \rangle -
\langle S_i ^z \rangle \langle S_j ^z \rangle \right]$, where
$S_i^z$ denotes the $z$-component of the spin at site $i$, we used
the density-matrix renormalization group (DMRG)
method.~\cite{White1992} We studied single chains (two coupled
chains) with length up to $L$ = 512 ($L$ = 96) and keeping up to $m$
= 4000 ($m$ = 2000) density-matrix eigenstates in the
renormalization procedure such that the truncation error was less
than 10$^{-9}$ (10$^{-6}$). In the single-chain case, the calculated
values were extrapolated to the thermodynamic limit $L \to \infty$.
We calculated the magnetization curve at very low temperature ($T/J$
$\ll$ 1) using the DMRG technique. The magnetization for a given
magnetic field was obtained as

\begin{equation}
M=\frac{\sum_{n=0}^{m_{\rm c}} \langle \psi_n | S_z | \psi_n \rangle
 \exp(-\frac{E_n}{k_{\rm B}T})}{\sum_{n=0}^{m_{\rm c}} \exp(-\frac{E_n}{k_{\rm B}T})},
\end{equation}

\noindent where $E_n$ and $\psi_n$ are the $n$-th eigenenergy and
the eigenstate ($n$ = 0 denotes the ground state), respectively,
$k_B$ is the Boltzmann constant, and $m_{\rm c}$ is a cutoff number
of the excited-state energies. The cutoff number $m_{\rm c}$
desirably should be sufficiently smaller than the number of states
$m$ kept in the density-matrix renormalization step. In this paper,
for a fixed system length $L$ = 64 and temperature $T$ = 1.8\,K,
$m_{\rm c}$ varied from 40 to 100 while keeping $m$ = 1000 and $M$
was extrapolated to the large $m_{\rm c}$ limit.

Applying the TMRG technique we analyzed the magnetic spin
susceptibility, $\chi(T)$, to obtain its maximum position,
$T^{max}_{\chi}$, as a function of the frustration parameter
$\alpha$. Notice that in the adopted random phase approximation
(RPA) for the interchain coupling (IC) $T^{\rm max}_{\chi}$ of the
3D susceptibility, $\chi_{3D}(T)$, is independent of the IC, the
$g$-factor, and the background susceptibility $\chi_0$.

\subsubsection{Local spin density L(S)DA+$U$ calculations}

In the second part, we calculate the total energies for various
prepared magnetic states, {\it i.e.}, a ferromagnetic, and various
antiferromagnetic states, whose total energy differences were mapped
onto those of corresponding spin states of a generalized $J_1$-$J_2$
model with supplemented interchain interactions. This way, the main
exchange integrals could be extracted.~\cite{schmitt09,koo11} The
density functional theory (DFT) based electronic-structure
calculations were performed using the full potential local orbital
scheme FPLO9.00-33.~\cite{fplo1,fplo2} Within the scalar
relativistic calculation the exchange and correlation potential of
Perdew and Wang was applied.~\cite{PW92} For the LSDA+$U$
calculations we varied $U$ in the physical relevant range from 5 to
8 \,eV using the mean-field approximation of the double-counting
correction. To ensure convergency we considered 518 $k$ points
within the irreducible part of the Brillouin zone.

\section{Magnetic Susceptibility}

\begin{figure}
\begin{center}
\includegraphics[width=0.95\columnwidth]{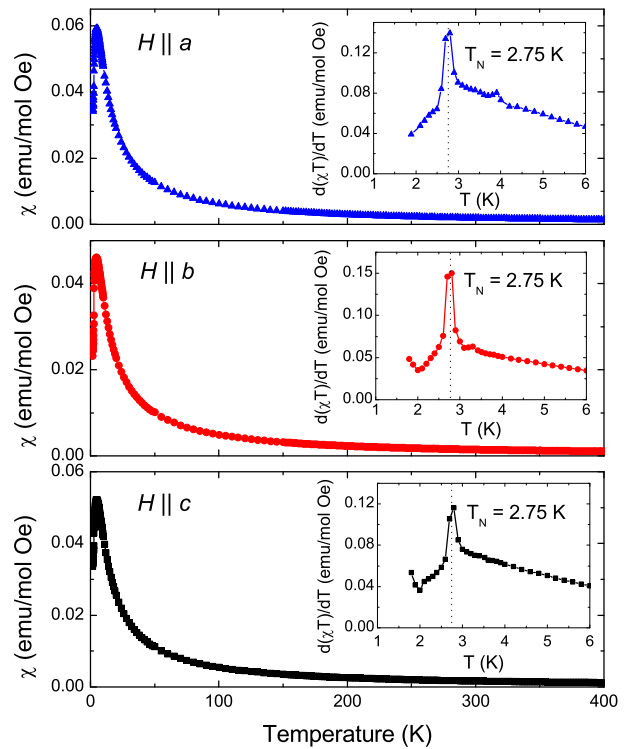}
\end{center}
\caption[1]{(Color online) Temperature dependence of the magnetic
susceptibility of PbCuSO$_4$(OH)$_2$ for an external magnetic field
of 0.4 T applied parallel to the $a$, $b$, and $c$ axes. In the
insets, the derivatives d($\chi$$T$)/d$T$ as function of temperature
are shown, from which the magnetic ordering temperature $T_{\rm N}$
$\approx$ 2.75(5)\,K can be evaluated.} \label{fig:sus}
\end{figure}

Figure~\ref{fig:sus} displays the temperature dependence of the
macroscopic susceptibility $\chi(T)$ of linarite with the magnetic
field applied along the three principle crystallographic directions.
The susceptibility exhibits two characteristic features in the
low-temperature region, {\it i.e.}, a maximum around $T^{max}_\chi$
= 4.9 $\pm$ 0.3\,K (averaged over all three crystallographic
directions, see below), just above the N\'eel temperature
$T_\mathrm{N}$ visible as a pronounced kink around 2.8\,K. While the
maximum around 5\,K is very common in quasi-1D frustrated
"ferromagnets" in the vicinity of the critical ferromagnetic-helical
point~\cite{drechsler07} (see also Fig.\ 13 in Sect.\ VI) and is
associated to low-lying ferromagnetic excitations, the kink at lower
temperatures is related to the transition into a long-range
magnetically ordered state. From the derivative d($\chi$$T$)/d$T$ of
the susceptibility data (insets in Fig. 2) the transition
temperature $T_\mathrm{N}$ = 2.75(5)\,K is determined.

While the qualitative behavior of $\chi(T)$ with respect to the
three crystallographic directions is the same, the absolute values
of $\chi_{max}$ at $T^{max}_\chi$ differ slightly with
$T^{max}_\chi$ = 5.0 $\pm$ 0.2\,K for $H$ $||$ $a$, $T^{max}_\chi$ =
4.6 $\pm$ 0.3\,K for $H$ $||$ $b$, and $T^{max}_\chi$ = 4.9 $\pm$
0.2\,K for $H$ $||$ $c$. From this observation, one already might
speculate that the $b$-axis should be the easy axis, despite of the
larger error bars of $T^{max}_\chi$. We would like to stress, that
our $T^{max}_\chi$ values are in agreement with those reported in
Ref.~\onlinecite{baran06}, while they differ from those of Yasui et
al.,~\cite{yasui11} where $T^{max}_\chi$ lies between 3.8 - 5\,K and
becomes maximal for $H$ $||$ $a'$. The origin of this discrepancy is
not clear up to now, although one possible scenario could be related
to small impurity contributions yielding an additional Curie-like
contribution, this way shifting $T^{max}_\chi$ to lower
temperatures.

\begin{figure}[b]
\begin{center}
\includegraphics[width=1\columnwidth]{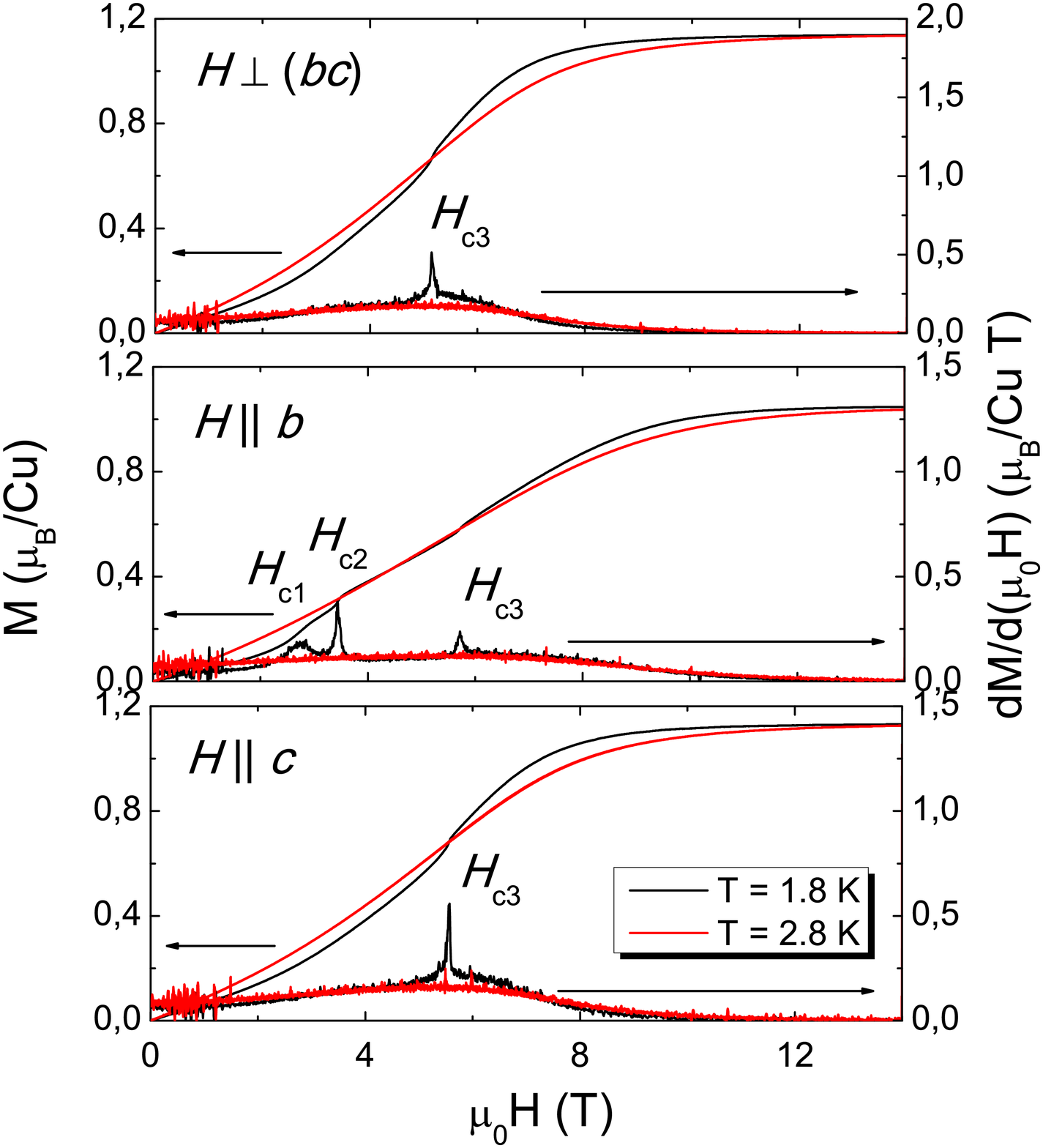}
\end{center}
\caption[1]{(Color online) Magnetization curve $M(\mu_0H)$ and its
derivative d$M$/d$(\mu_0H)$ for magnetic fields applied along
$a_{\perp}$, $b$, and $c$ of PbCuSO$_4$(OH)$_2$ measured at $T$ =
1.8 K $< T_\mathrm{N}$ (black curves) and $T$ = 2.8 K $\geq
T_\mathrm{N}$ (red curves).} \label{fig:magnetization}
\end{figure}

In Fig.~\ref{fig:magnetization}, we present the magnetization data
$M(\mu_0H)$ and the derivatives d$M$/d($\mu_0H$) of
PbCuSO$_4$(OH)$_2$ as a function of field at $T$ = 1.8 K $<$
$T_\mathrm{N}$ and $T$ = 2.8\,K $\geq$ $T_\mathrm{N}$ with $H$
$\perp$ (bc) (hereafter named $a_{\perp}$), $H$ $||$ $b$, and $H$
$||$ $c$.~\cite{note1} Although here we will focus on the properties
of linarite in the paramagnetic regime, we added the low-temperature
data in order to extract the saturation field, $H_s$, in the case of
reduced thermal fluctuations. Measurements were made for increasing
and decreasing field; no hysteresis has been observed within the
experimental resolution as result of a low sweep rate. Comparing the
magnetization versus field (and its derivative), a large anisotropic
response is observed with $M_{s,a_{\perp}}$ $\approx$ 1.16
$\mu_\mathrm{B}$/Cu atom, $M_{s,b}$ $\approx$ 1.05
$\mu_\mathrm{B}$/Cu atom, and $M_{s,c}$ $\approx$ 1.15
$\mu_\mathrm{B}$/Cu atom for the $a_{\perp}$, $b$, and $c$
direction, respectively. Together with the anisotropic values of the
saturation field $\mu_0H_s$, {\it i.e.}, $\mu_0H_{s,a_{\perp}}$
$\approx$ 7.6\,T, $\mu_0H_{s,b}$ $\approx$ 10.5\,T, and
$\mu_0H_{s,c}$ $\approx$ 8.5\,T, this implies an appreciable
anisotropy of the magnetic exchange in our compound.

The anisotropic magnetic behavior of PbCuSO$_4$(OH)$_2$ is also
evidenced in the number of transitions observed in the
$dM/d(\mu_0H)$ curve at 1.8\,K. While for the $a_\perp$ and $c$
directions only one sharp peak indicative of a phase transition is
observed (around $\mu_0H_{c3}$ $\approx$ 5.2-5.5 T), for the $b$
direction, {\it i.e.}, for the chain direction, three clear
transitions could be resolved at $\mu_0H_{c1}^{b}$ $\approx$ 2.7 T,
$\mu_0H_{c2}^{b}$ $\approx$ 3.3 T, and $\mu_0H_{c3}^{b}$ $\approx$
5.8\,T. These transitions at low fields possibly assign rotation and
spin-flop reorientations out of a helical ground state for the field
aligned along the easy magnetic axis. We will discuss this behavior
in more detail in a forthcoming paper.~\cite{willenberg11}

Both the anisotropy and the overall small saturation field of
linarite are very important and unique features with respect to the
predicted exotic high-field phases. This makes PbCuSO$_4$(OH)$_2$
suitable for the investigation of such phenomena in easily available
static magnetic fields. In order to resolve the origin of the
magnetic anisotropy in linarite, which could either stem from
\textit{g}-factor anisotropy, anisotropic exchange, or antisymmetric
Dzyaloshinskii-Moriya interactions to name a few, we performed ESR
investigations.

\section{Electron Spin Resonance (ESR)}

\begin{figure}[t]
\begin{center}
\includegraphics[width=0.9\columnwidth]{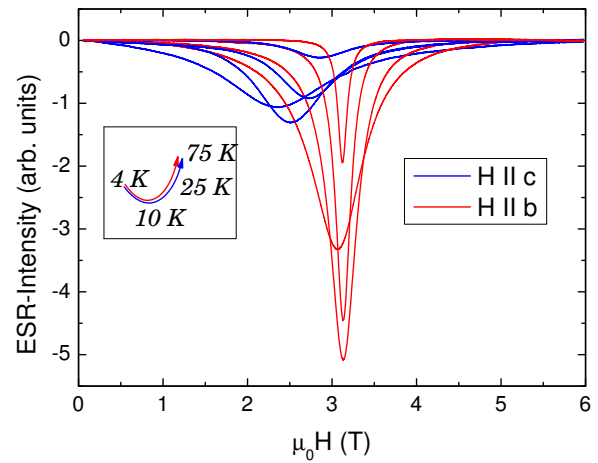}
\end{center}
\caption[1]{(Color online) ESR spectra for different temperatures
measured at 93\,GHz with the magnetic field $H$ applied along the
\emph{b} and \emph{c} axes. Spectra for $H$ $||$ \emph{a} (not
shown) are similar to spectra for $H$ $||$ $c$. Arrows on the left
side indicate temperatures of the individual ESR spectra.}
  \label{ESR-spectra}
\end{figure}

ESR measurements were performed for different single crystals of
linarite. A small single crystal from \emph{origin 2} was used for
the measurements in the resonant cavity at a frequency of about
93\,GHz. At this frequency, the resonance field is found around
3\,T. The spectrum consists of a single line of Lorentzian shape
(Fig.~\ref{ESR-spectra}). From a fit to the ESR lines, the
intensity, the resonance field, and the linewidth are extracted.
From those parameters the integrated ESR intensity, which is
determined by the intrinsic spin susceptibility, was calculated (see
Fig.~\ref{theta-CW} and discussion in the following NMR section).
Since the resonator was inserted into the cryostat its quality
factor $Q$ and its resonance frequency $\nu_{\mathrm{res}}$ change
with temperature. $Q$ and $\nu_{\mathrm{res}}$ are determined with
frequency sweeps around the resonance frequency at every temperature
and the spin susceptibility $\chi_{s}$ and resonance field $H_{res}$
can be corrected accordingly.

\begin{figure}[t]
\begin{center}
\includegraphics[width=0.8\columnwidth]{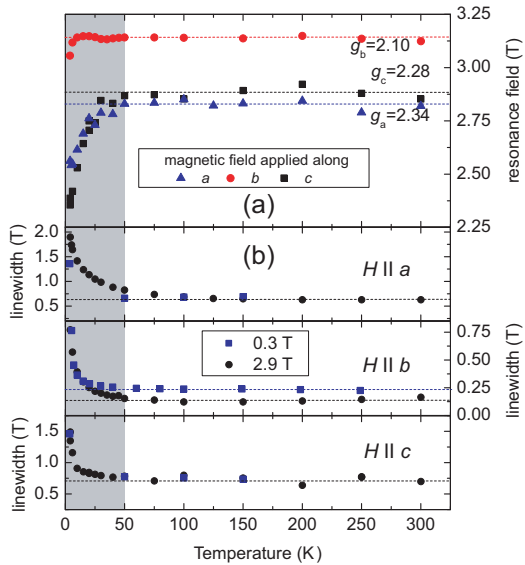}
\end{center}
  \caption[1]{(Color online) (a) Resonance fields at about 93\,GHz and
  (b) linewidths for resonance fields of about 0.3 and 2.9\,T of the
  ESR signals as a function of temperature for the external magnetic
  field applied along the crystallographic axes \emph{a}, \emph{b}, and
  \emph{c}. The resonance field is temperature independent above
  $\sim$ 50\,K and decreases with decreasing temperature. For the
  high-temperature resonant fields the corresponding $g$-factors are
  listed. The linewidth is significantly smaller for $H$ $||$ $b$.}
  \label{g-factors}
\end{figure}

The temperature dependencies of the ESR resonance field and ESR
linewidth are plotted in Fig.~\ref{g-factors}.~\cite{noteESR} In the
high-temperature regime (roughly above 50\,K) the resonance fields
are temperature independent. A shift of the resonance position of an
ESR signal as a function of temperature is associated with the
development of internal magnetic fields in the system. Along the
\emph{b} direction a shift is observed only at low temperatures
close to the ordering temperature $T_\mathrm{N}$. This shows that an
internal field develops along the chain direction only when the
actual 3D ordering occurs. However, along the \emph{a} and \emph{c}
directions a shift in the resonance field is observed already at
much higher temperatures starting at around 50\,K and developing
smoothly with decreasing temperature. This indicates that the Cu
spins are primarily aligned in the \emph{ac} plane, but do not point
along the chain direction in the paramagnetic regime. The internal
fields developing are, therefore, predominantly directed
perpendicular to the chain direction. At first glance, the above
considerations are in some conflict with the magnetization
measurements indicating that the $b$ axis might be the easy axis of
our system. However, taking into account that the magnetic field
used in the ESR measurements ($\sim$ 3 Tesla) will probably be
strong enough to rotate the pre-spiral orientation out of the
easy-plane (see Sect. III), even at high temperatures above
$T_\mathrm{N}$ short-range ordered clusters could explain our
observations with predominant internal fields perpendicular to the
chain direction.

From the resonance field, $H_{res}$, the effective $g$-factors along
the crystallographic directions can be determined as $g=h\cdot
\nu_{res}/(\mu_{\mathrm{B}}H_{res})$. For the high-temperature
regime the effective $g$-factors are found to be $g_{a}=2.34$,
$g_{b}=2.10$, and $g_{c}=2.28$.

\begin{figure}[b]
\begin{center}
\includegraphics[width=0.9\columnwidth]{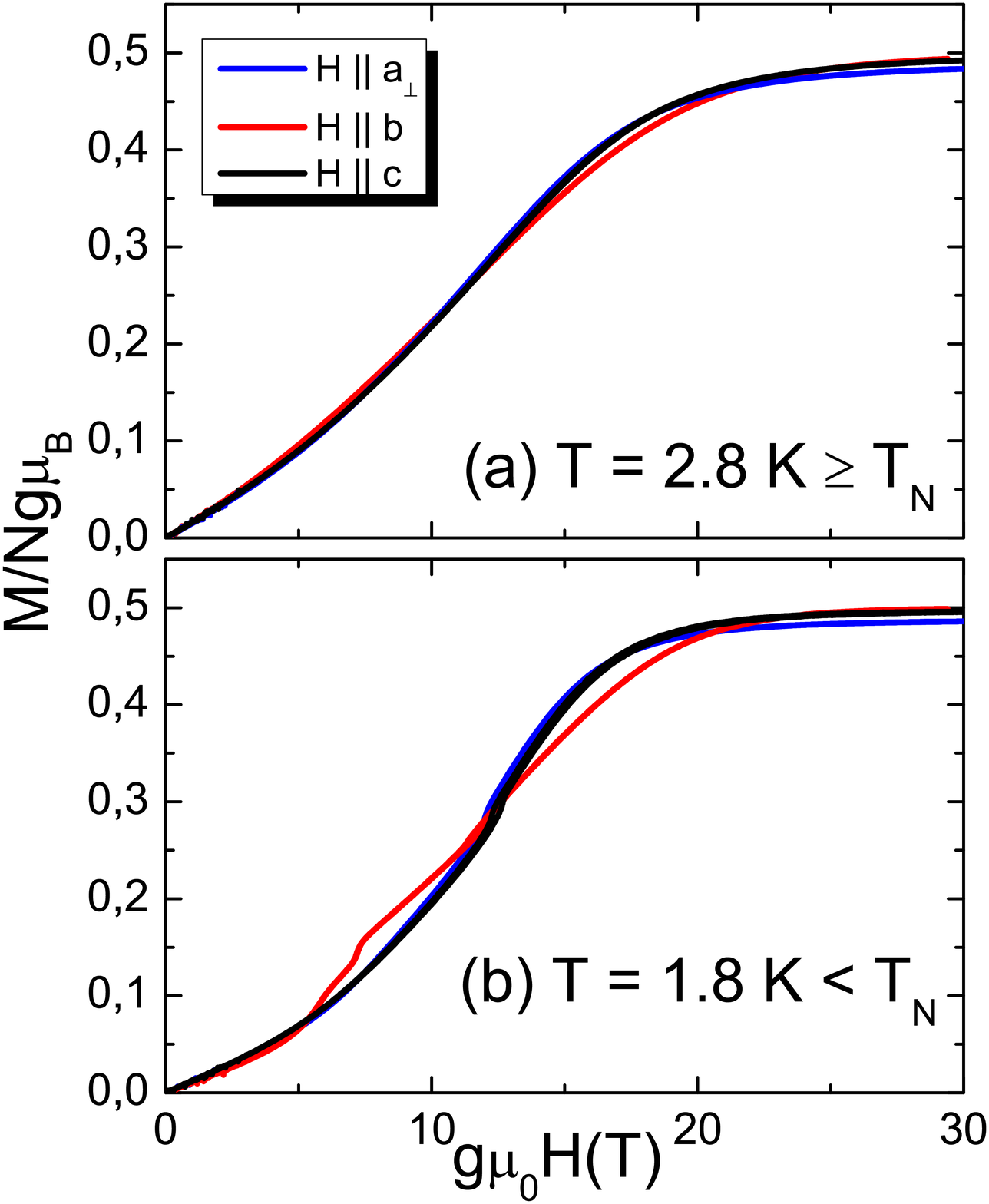}
\end{center}
  \caption[1]{(Color online) The experimentally determined spin expectation
value ${<S_z>}$ = $M/Ng\mu_\mathrm{B}$ of PbCuSO$_4$(OH)$_2$ as a
function of the scaled magnetic field $g\mu_0H$ along different
crystallographic directions and temperatures (a) $T$ = 2.8 K $\geq$
$T_\mathrm{N}$ and (b) $T$ = 1.8 K $<$ $T_\mathrm{N}$.}
  \label{gscaled-magnetization}
\end{figure}

Having established the $g$-factors for the three principal
crystallographic directions, we can analyze the magnetic anisotropy
in our system. In Fig.~\ref{gscaled-magnetization}, we present the
spin expectation value $<S_z> = M/Ng\mu_B$ (N: Avogadro number) of
PbCuSO$_4$(OH)$_2$ as a function of the scaled field $g\mu_0H$ at
2.8 and 1.8\,K and for the three crystallographic directions
$a_{\perp}$, $b$, and $c$ as derived from the experimentally
determined magnetization data $M(H)$ (Fig. 3). The extracted spin
expectation value corresponds to the Cu spin 1/2. In the
paramagnetic regime above $T_\mathrm{N}$, the anisotropy of both
saturation magnetization and saturation field is explained mainly by
the anisotropy of the $g$-factor. Note that there is a difference in
the calculated and directly measured saturation magnetization for
$H\parallel a_\perp$ due to the $g$-factor. For the ESR experiment
the magnetic field was aligned along $a$, while for the
magnetization measurement the field was aligned perpendicular to the
$bc$ plane. For an alignment along $a$ the saturation magnetization
would be slightly larger and would match the calculated value. For
temperatures smaller than $T_\mathrm{N}$, however, the anisotropy
cannot be described by the $g$-factor anisotropy. Additional
contributions from symmetric exchange anisotropy and/or possibly
Dzyaloshinskii-Moriya interactions need to be taken into account.

As shown in Fig.~\ref{g-factors}, the ESR linewidth is strongly
anisotropic. At a frequency of 93\,GHz corresponding to a resonance
field of about $H_{res} \approx 3$\,T, the linewidth for $H
\parallel b$ is with $\Delta H \approx
0.13$\,T much smaller than $\Delta H \approx 0.7$\,T for $H
\parallel a$ and $c$. For all directions the linewidth is almost
constant or only weakly dependent on temperature above 50\,K. At the
X-band frequency of about 9.6\,GHz - corresponding to $H_{res}
\approx 0.3$\,T - the temperature dependencies of the ESR signals
show a similar behavior (squares in Fig.~\ref{g-factors}(b)). The
linewidths along \emph{a} and \emph{c} are - within the uncertainty
of the measurement - identical to the linewidths at higher fields.
The decreasing intensity of the ESR signal with increasing
temperature makes it difficult to analyze the lines perfectly along
those orientations up to room temperature, however, the linewidth
appears to stay constant. For the \emph{b} direction at high
temperatures, the linewidth is constant and fairly narrow with
$\Delta H\approx0.25$\,T, which is unexpected since it is broader
than for the larger field of about 3\,T. This effect can be
explained by the strongly anisotropic linewidth together with a
slight misalignment of the sample of about 5$^\circ$ for this
particular measurement.

On approaching temperatures below 50\,K a broadening of the lines is
observed for both fields. The small resonance field of about 0.3\,T
limits the reliability of the measurements for fields aligned along
\emph{a} and \emph{c}, since the linewidth exceeds the resonance
field and a fit to the data cannot be done accurately anymore.
However, the values are close to the linewidths for the ten times
larger field of about 3\,T.

\begin{figure}[b]
\begin{center}
\includegraphics[width=0.9\columnwidth]{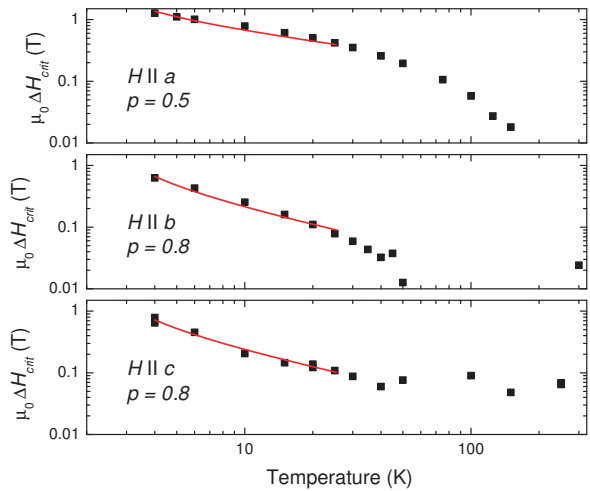}
\end{center}
\caption[1]{(Color online) Double logarithmic plot of the critical
linewidth part $\Delta H_{crit}$ (from ESR) as a function of
temperature for $H$ parallel to the three crystallographic
directions \emph{a}, \emph{b}, and \emph{c}. Fits to the data for
temperatures up to 25\,K are shown.}
 \label{loglog-width}
\end{figure}

The ESR linewidth depends on (dipolar) spin-spin or anisotropic
exchange interactions as well as on the development of internal
magnetic fields. The fact that the linewidth is the same for
different fields indicates that inhomogeneous broadening effects are
rather small. As we approach lower temperatures the spin-spin
correlation length increases and short-range magnetic correlations
develop. The change of linewidth as a function of temperature yields
information about the dimensionality and type of interactions in the
system. The broadening can be analyzed in terms of

\begin{equation}
\Delta H (T) = \Delta H_0+C[(T-T_{\mathrm{N}})/T_{\mathrm{N}}]^{-p}
\end{equation}

\noindent as the ordering temperature is approached. The linewidth,
$\Delta H (T)$, is divided into a non-critical constant part,
$\Delta H_0$, and a temperature-dependent critical part, $\Delta
H_{crit} (T)$.~\cite{Jongh90} The exponent $p$ yields information
about the effective dimensionality of the correlated spin system and
its change by approaching a long-range-ordered ground state. Fits to
our data over the low-temperature range with a fixed $T_{\mathrm{N}}
= 2.0$\,K at 3\,T result in critical exponents $p$ = 0.5-0.8
(Fig.~\ref{loglog-width}). Note, that the magnetic ordering
temperature $T_\mathrm{N}$ = 2.0 \,K was chosen according to the
field dependence of $T_\mathrm{N}$ as determined by thermodynamic
bulk measurements.~\cite{willenberg11} For a 1D Heisenberg magnet
the critical exponent was found to be about $p$ =
2.5.~\cite{Tazuke75} However, on approaching the ordering
temperature this value can change significantly and values of $p$
$\approx$ 0.6 have been reported.~\cite{Jongh90} Such a critical
exponent is interpreted as signalling the appearance of 3D
antiferromagnetic fluctuations. The fact that the broadening in our
system occurs already at temperatures around 50\,K, {\it i.e.}, 15
times higher than the actual ordering temperature, points to
appreciable magnetic fluctuations at elevated temperatures
indicative of substantial interlayer correlations well above
$T_\mathrm{N}$. The actual 3D ordering at much lower temperatures
then indicates a strongly frustrated system with competing
interactions on the energy scale of $\sim$ 50\,K.

\section{Nuclear Magnetic Resonance (NMR)}

Our NMR experiments were conducted using $^1$H- and $^{207}$Pb as
probing nuclei. Both the hydrogen and the lead ions occupy
low-symmetric crystallographic sites with respect to the magnetic Cu
sites, {\it i.e.}, none of the ions are located exactly between two
Cu ions neither along the chain nor between two neighboring chain
structures. While there is only one crystallographic site for Pb,
two inequivalent H sites can be expected according to recent
structural investigations via neutron diffraction emphasizing two
different kinds of hydrogen bondings in linarite.~\cite{schofield09}
Henceforth, a single $^{207}$Pb and two $^1$H-NMR lines can
generally be expected in our experiment. We name the hydrogen site
with the stronger bonding H$_1$, {\it i.e.}, probably the one with
the larger hyperfine coupling, and the one with the weaker bonding
H$_2$. For magnetic fields $H$ $||$ $b$, however, two nearly
overlapping $^1$H resonance lines imply that both hyperfine
couplings are nearly identical for this field direction. We have
also tried to detect the $^{63,65}$Cu spin-echo signals, but did not
succeed. We attribute the lack of copper signal to very short
spin-spin relaxation times $T_2$ of linarite, which are of the order
of 20 $\mu$s at room temperature even at the $^{207}$Pb and $^1$H
sites.

The interest in probing both $^1$H and $^{207}$Pb lies in the
different coupling of the two nuclei with their neigboring atoms and
thus different distances and symmetries with respect to the magnetic
Cu$^{2+}$ ions. From the crystallographic structure and
chemical-bonding scheme it can be expected that due to the distance
between Pb and neighboring magnetic Cu ions the hyperfine coupling
at the $^{207}$Pb site will be dominated by dipolar couplings
between Pb nuclei and Cu spins. At the $^{207}$Pb site, the dipolar
fields will be predominantly given by the spins of the two nearest
magnetic ions, {\it i.e.}, two Cu spins along the chain ($b$)
direction. But also couplings to next-nearest neighboring Cu ions
along the chain as well as neighboring chains along $c$ and $a$ can
be expected to result in small additional contributions. On the
other hand, at the $^1$H site, the effective local fields at the
probing nuclei are probably composed of the dipole fields of
surrounding magnetic Cu$^{2+}$ moments and of so-called contact
fields. The latter are due to the direct neighboring environment of
hydrogen and oxygen atoms, which mediate the magnetic superexchange
between magnetic Cu ions. In this situation a small polarization of
the hydrogen atoms can be envisaged. Since the hydrogen atoms are
located very close to the $bc$ plane of Cu$^{2+}$ ions, the
hyperfine fields will be predominantly given by the spins of the
four nearest magnetic ions of the nearest neighboring $bc$ plane,
{\it i.e.}, two neighboring Cu spins from one and two neighboring Cu
spins from a second Cu-chain shifted by the lattice constant $c$.

\begin{figure}
\begin{center}
\includegraphics[width=1\columnwidth]{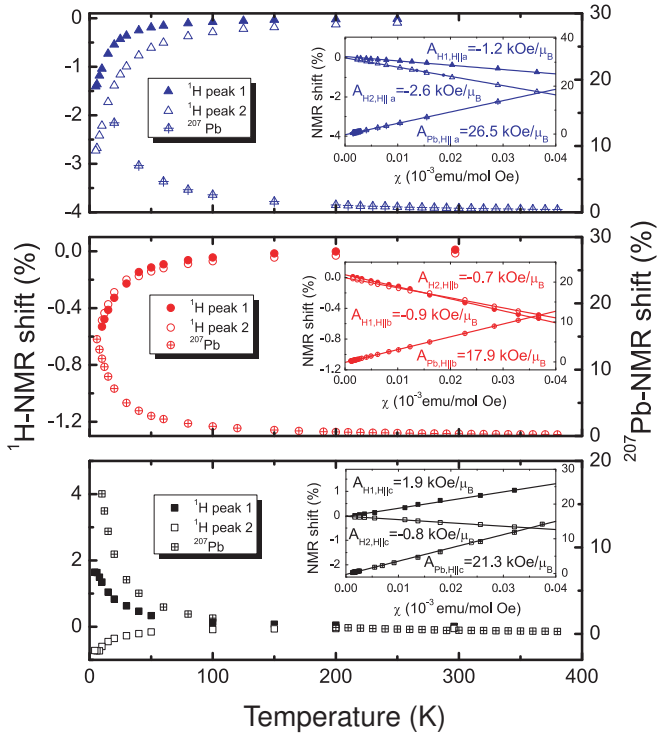}
\end{center}
  \caption[1]{(Color online) The $^1$H- and $^{207}$Pb-NMR shift of
  PbCuSO$_4$(OH)$_2$ in the temperature range 5 - 400\,K for all three
  crystallographic directions. While the $^1$H data have been determined
  in an external field of 2\,T, the $^{207}$Pb-NMR shift has
  been measured in an applied field of 4\,T. In the insets, the
  NMR shifts are plotted as function of the macroscopic bulk susceptibility.
  The lines represent linear fits to the experimental data; for details see text.}
  \label{fig:knightshift}
\end{figure}

The spin-echo signal of the $^1$H and $^{207}$Pb lines was observed
in the temperature range 5 - 400\,K. We first analyze the resonance
shift of the $^1$H and $^{207}$Pb-NMR lines for an applied magnetic
field of 2 and 4\,T, respectively (Fig.~\ref{fig:knightshift}). The
different field values have been used in order to obtain reasonable
frequency ranges for both nuclei. Due to strong transverse magnetic
short-range correlations leading to a very short spin-spin
relaxation time $T_2$ of less than 5 $\mu$s at low temperatures, a
wipe-out of the $^{207}$Pb-NMR signal occurs below $\sim$ 10\,K. The
NMR shift is defined as the normalized difference between the
observed resonance frequency, $\omega_{res}$, and the calculated
value for the bare nucleus,

\begin{equation}
K (T) = \frac{\omega_{res}-\gamma\mu_0H_0}{\gamma\mu_0H_0},
\label{eq:nmrshift}
\end{equation}

\noindent $\gamma$ being the gyromagnetic ratio of the nucleus and
$H_0$ being determined from the $^1$H and $^2$D-NMR resonance
frequency of (deuterated) water at room temperature. From
Fig.~\ref{fig:knightshift} a strong paramagnetic increase of both
the $^1$H- and $^{207}$Pb-NMR shift is observed with decreasing
temperature, which arises from the interactions between the probing
nuclei and the surrounding electrons.

Generally, the NMR shift $K_{tot}(T)$ can be divided into two
contributions,

\begin{equation}
K_{tot} (T) = K_{spin}(T) + K_{orb}, \label{eq:KspinKorb}
\end{equation}

\noindent where $K_{spin}(T) = A \chi_{spin}(T)$ arises via a
hyperfine coupling to the electronic spins and $K_{orb}$ stems from
a temperature-independent orbital magnetization induced at the
nucleus site. Here, $A$ is the hyperfine coupling constant, which
can either have a positive or negative sign, leading to positive or
negative temperature dependencies of the NMR shift. From this
equation it is already obvious that NMR has an advantage over bulk
susceptibility investigations. Via NMR one accurately measures the
intrinsic spin susceptibility, $\chi_{spin}(T)$, without suffering
from temperature-independent diamagnetic core or Van Vleck
contributions, from free spins (impurities) and extrinsic foreign
phases, which limits the accuracy of bulk susceptibility
measurements. Therefore, it is more reliable to extract the magnetic
parameters from the temperature dependence of the NMR shift rather
than from bulk susceptibility. The conventional scheme of
correlating the NMR shift $K_{tot}(T)$ and the bulk susceptibility
$\chi(T)$ is to plot both parameters as $K_{tot}(\chi)$ with
temperature being an implicit parameter. This way, the slope yields
the hyperfine coupling constant $A$, while $K_{orb}$ results from
the intersect at $\chi$ = 0.

The NMR shift as function of the bulk susceptibility of linarite is
shown in the insets of Fig.~\ref{fig:knightshift} for both nuclei
and for the magnetic field applied along the three crystallographic
directions. Clearly, both physical properties scale with each other
for all cases in the full temperature regime. A linear fit to this
data yields highly anisotropic hyperfine coupling constants for both
$^1$H-nuclei, {\it i.e.}, $A_{H_1,H||a}$ = -1.2 kOe/$\mu_\mathrm{B}$
and $A_{H_2,H||a}$ = -2.6 kOe/$\mu_\mathrm{B}$, $A_{H_1,H||b}$ =
-0.9 kOe/$\mu_\mathrm{B}$, $A_{H_2,H||b}$ = -0.7
kOe/$\mu_\mathrm{B}$, $A_{H_1,H||c}$ = 1.9 kOe/$\mu_\mathrm{B}$ and
$A_{H_2,H||c}$ = -0.8 kOe/$\mu_\mathrm{B}$. The different hyperfine
couplings of the two inequivalent H atoms strongly support the
notion of different hydrogen bondings to neighboring oxygen sites in
linarite as determined previously.~\cite{schofield09} In contrast to
these anisotropic values for the $^1$H-nuclei, the $^{207}$Pb
hyperfine couplings are dominated by a large positive isotropic
contribution, which is complemented by a small anisotropic (dipolar)
component for the three different axes, yielding overall large and
positive values of $A_{Pb,H||a}$ = 26.5 kOe/$\mu_\mathrm{B}$,
$A_{Pb,H||b}$ = 17.9 kOe/$\mu_\mathrm{B}$, and $A_{Pb,H||c}$ = 21.3
kOe/$\mu_\mathrm{B}$ for fields aligned along $a$, $b$, and $c$,
respectively.

\begin{figure}[b]
\begin{center}
\includegraphics[width=0.85\columnwidth]{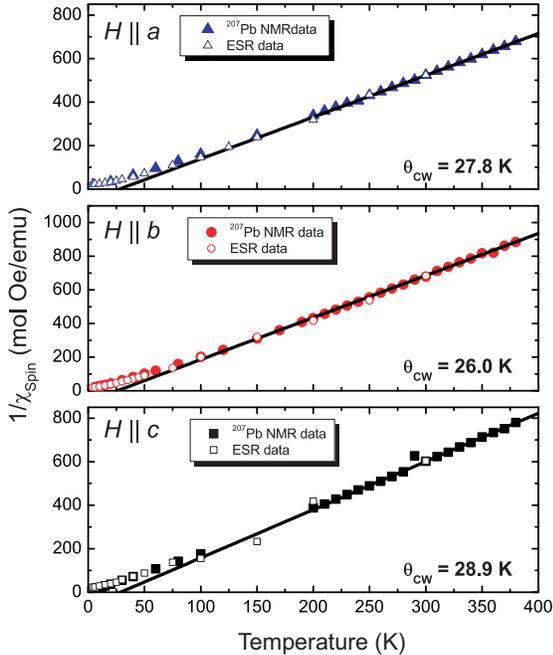}
\end{center}
  \caption[1]{(Color online) Inverse spin susceptibility of linarite,
  $\chi_{spin}^{-1}(T)$, for the external magnetic field applied along
  \emph{a}, \emph{b}, and \emph{c} as determined via $^{207}$Pb-NMR and
  ESR. The spin susceptibility as determined via ESR at about 3\,T was
  normalized to the high-temperature (300\,K) value of the static
  susceptibility. The lines represent linear fits to a Curie-Weiss law
  in the high-temperature range from 250 to 400\,K.}
  \label{theta-CW}
\end{figure}

After having determined the hyperfine couplings for the two
different nuclei, the intrinsic spin susceptibility of
PbCuSO$_4$(OH)$_2$ can be evaluated via $\chi_{spin}(T) =
(K_{tot}(T) - K_{orb})/A$ for the three crystallographic directions.
Fig.~\ref{theta-CW} depicts these physical properties plotted as the
inverse spin susceptibility, $\chi_{spin}^{-1}$, as a function of
temperature as derived from the $^{207}$Pb-NMR data.~\cite{note2}
For a comparison we added the spin susceptibility obtained from our
ESR investigations after normalizing these data to the value of the
static susceptibility at 300\,K. From this figure it is clearly
visible that the intrinsic spin susceptibility from ESR and NMR
scales nicely and is practically identical over the whole
temperature range for all crystallographic directions.

Then, from a linear fit of the inverse susceptibility to a
Curie-Weiss law $\chi_{spin}^{-1}(T)$ $\propto$ ($T$ -
$\Theta_\mathrm{CW}$) in the $T$ region 250 - 400\,K, the
Curie-Weiss temperature, $\Theta_\mathrm{CW}$, could be determined
for the three directions. Its absolute value is isotropic within the
experimental error bars as expected for a Cu$^{2+}$ $s$ = 1/2
system, yielding $\Theta_\mathrm{CW}$ = 27(2)\,K. The positive value
of the Curie-Weiss temperature indicates the predominance of a
ferromagnetic coupling. Comparing the Curie-Weiss temperature to the
ordering temperature, a ratio $\Theta_\mathrm{CW}/T_\mathrm{N}$
$\sim$ 10 is extracted for linarite. This quantity is commonly used
to judge the level of frustration in a compound, since frustration
tends to suppress long-range order. A ratio of 10 shows that
frustration is definitely an important issue that needs to be
considered in this compound.

\begin{figure}
\begin{center}
\includegraphics[width=0.95\columnwidth]{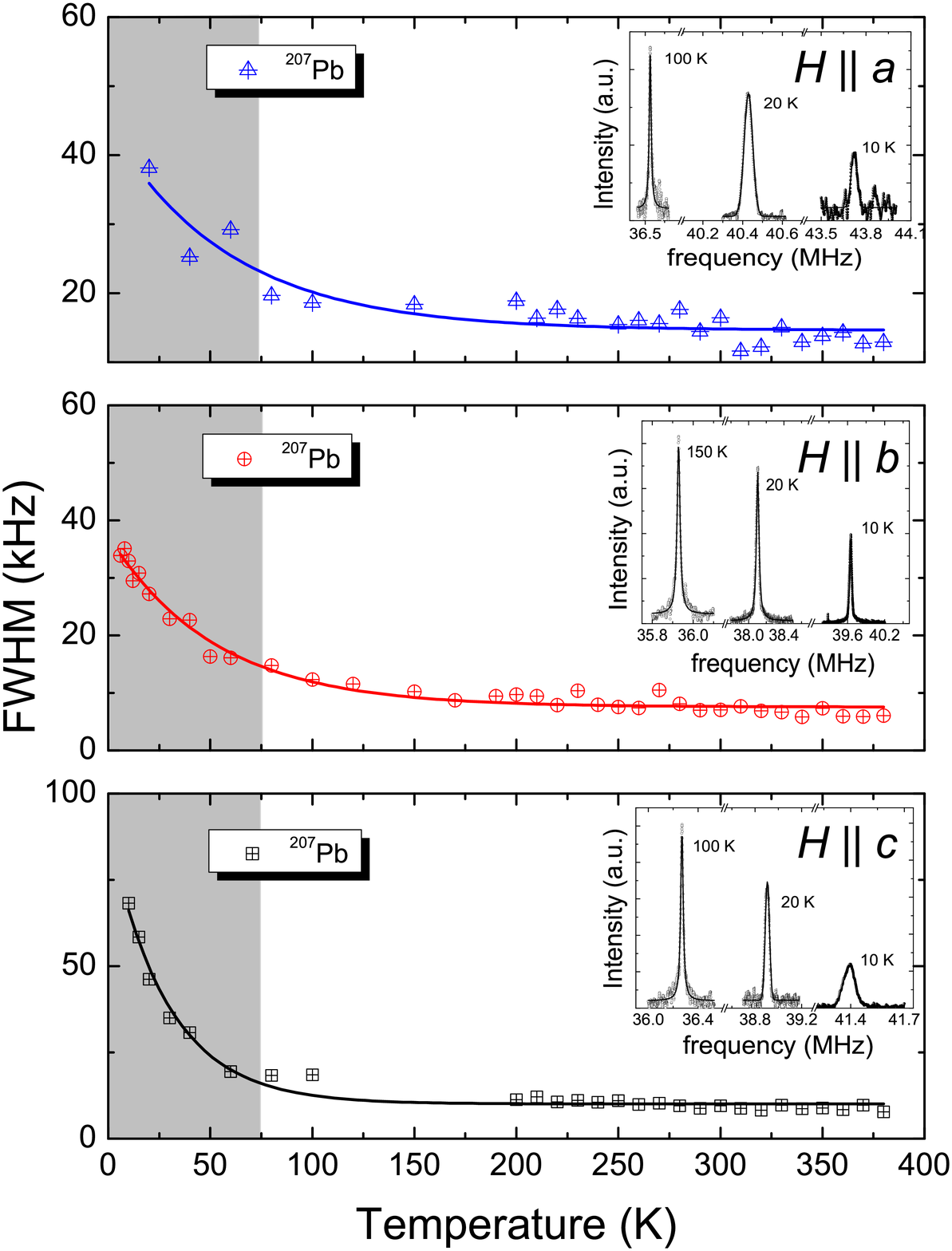}
\end{center}
  \caption[1]{(Color online) The $^{207}$Pb-NMR linewidth of
  PbCuSO$_4$(OH)$_2$ as a function of temperature in an external magnetic
  field of 4\,T applied along the three crystallographic axes $a$, $b$,
  and $c$. The lines are guides to the eyes. The insets show representative
  $^{207}$Pb-NMR spectra for different temperatures. The different
  intensities at these temperatures are not to scale; for details see text.}
  \label{FWHM-Pb}
\end{figure}

\begin{figure}
\begin{center}
\includegraphics[width=0.95\columnwidth]{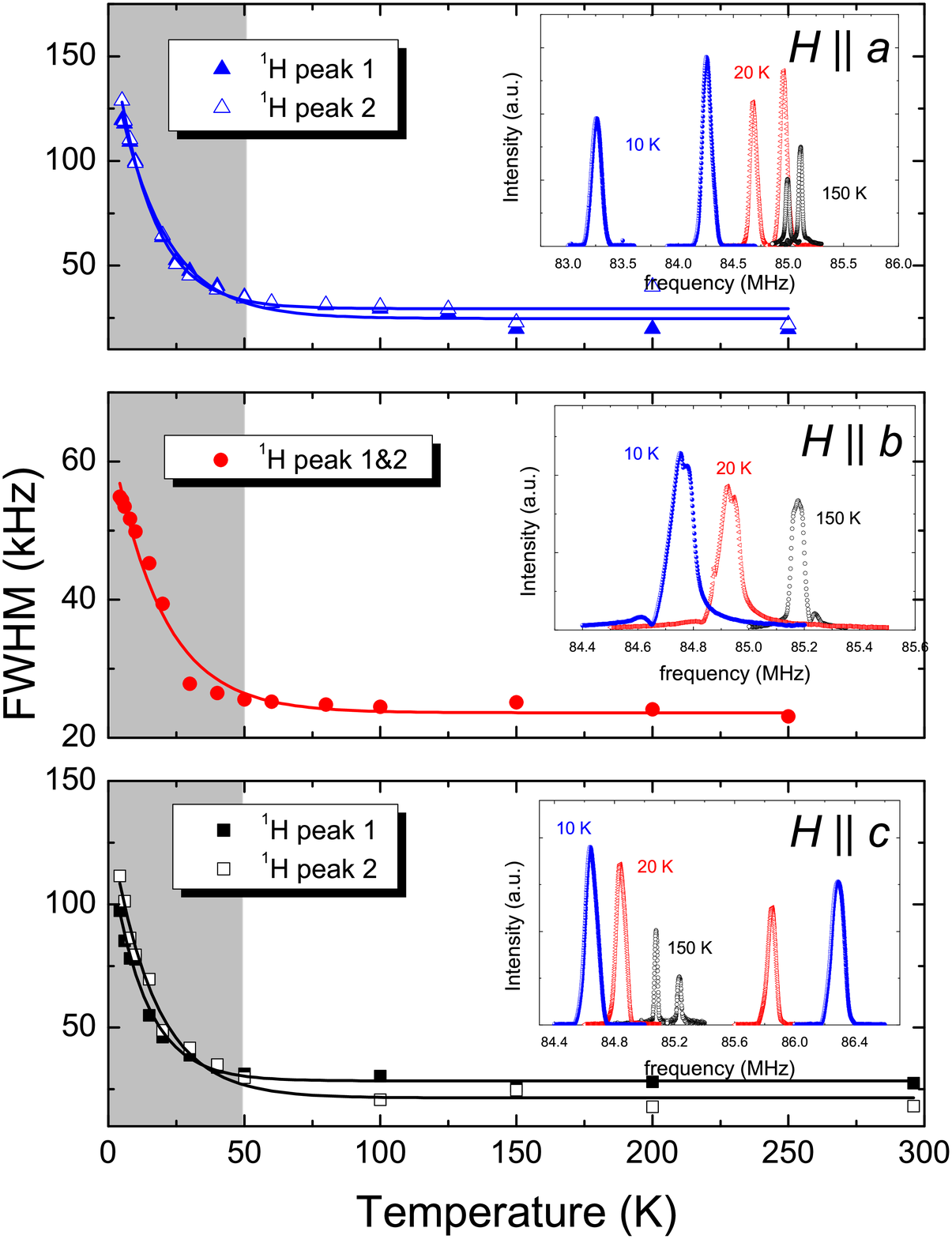}
\end{center}
  \caption[1]{(Color online) The $^1$H-NMR linewidth of
  PbCuSO$_4$(OH)$_2$ as function of temperature in an external magnetic
  field of 2\,T parallel to the three crystallographic axes $a$, $b$, and $c$.
  The lines represent guides to the eyes. The insets show representative
  $^1$H-NMR spectra for 10, 20, and 150\,K. The intensities at
  different temperatures are not to scale; for details see text.}
  \label{FWHM-H}
\end{figure}

In Figs.~\ref{FWHM-Pb} and ~\ref{FWHM-H}, the linewidths of the
$^{207}$Pb- and the $^1$H-NMR spectra, respectively, are shown as
function of temperature for all three crystallographic directions
$a$, $b$, and $c$. In the insets, some $^{207}$Pb and $^1$H-NMR
spectra are depicted for applied magnetic fields of 4 and 2\,T,
respectively, and for 3 different temperatures between 10 and
150\,K. In the paramagnetic phase, the spin-echo NMR signal has a
rather isotropic Lorentzian lineshape. Since the linewidth at high
temperatures is very small, {\it i.e.}, about 10-25 kHz for all
spectra and for all three crystallographic directions, we suggest
that our single crystal is of rather high-quality. Note that also
for the $^1$H-NMR spectra for $H$ $||$ $b$ two Lorentzian lines have
been used to fit the data since the two NMR lines do not perfectly
overlap at low temperatures. In Fig.~\ref{FWHM-H}, however, the
average of the linewidth of both $^1$H-spectra has been plotted for
$H$ $||$ $b$.

Similar to the ESR linewidth, the temperature dependence of the full
width at half maximum (FWHM) of the NMR spectra is expected to give
access to the dynamics of the magnetic correlations and thus to the
dynamical critical properties in the paramagnetic regime upon
approaching $T_\mathrm{N}$. The NMR linewidth is related to the
spin-spin relaxation time, $T_2$, and thus probes the transverse
component of the two-spin correlation function and the temporal spin
fluctuations of the magnetic system near the critical temperature.
Considering a compound with anisotropic magnetic exchange
interactions such as linarite, it can be expected that the NMR
linewidth is dominated by spin fluctuations along the magnetic easy
axis, with spin fluctuations perpendicular to the easy axis only
contributing to the non-critical broadening. Hence, taking into
account that the linewidth probes transverse spin fluctuations the
broadening of the NMR line should be most prominent for magnetic
fields perpendicular to this (easy) axis.

For all NMR spectra a pronounced broadening of the line has been
observed below $\sim$ 75\,K for the $^{207}$Pb signal and below
$\sim$ 50\,K for the $^1$H spectra. This broadening points to
short-range correlations developing already at temperatures $T$
$\gg$ $T_\mathrm{N}$. Comparing both the response at the two
different nuclei and for the three different crystallographic
directions, one can easily see that (i) the broadening of the NMR
line is shifted to lower temperatures but slightly enhanced for the
$^1$H spectra and (ii) that particularly for the 2\,T $^1$H-NMR data
the broadening is more pronounced for the directions perpendicular
to the Cu chain. The latter is in perfect agreement with the results
obtained from the temperature dependence of the ESR linewidth and
emphasizes the magnetic anisotropy in our system.

\begin{figure}[b]
\begin{center}
\includegraphics[width=0.85\columnwidth]{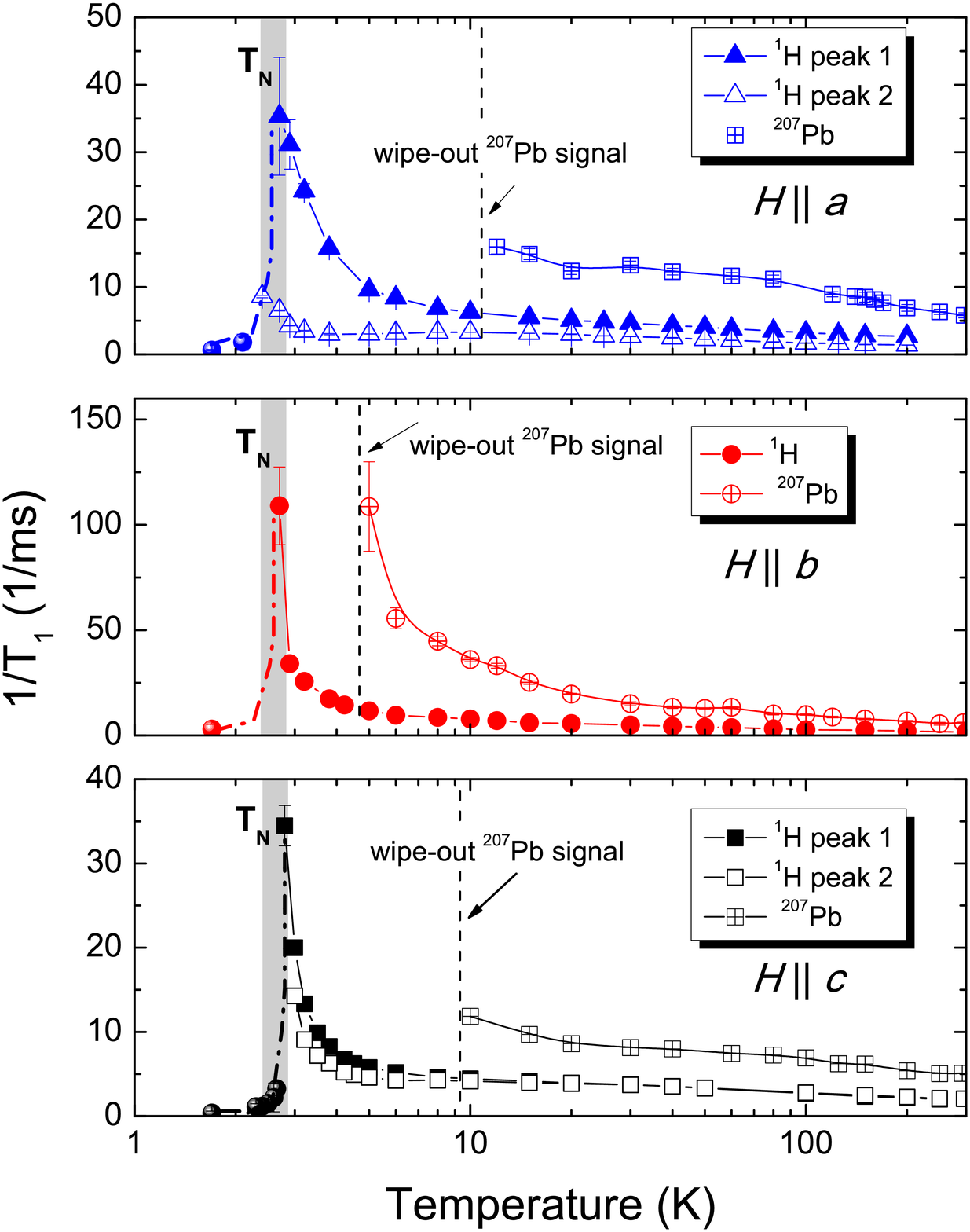}
\end{center}
  \caption[1]{(Color online) The spin-lattice relaxation rate, 1/$T_1$, of
  PbCuSO$_4$(OH)$_2$ as function of temperature determined in an external
  field of 2 and 4\,T for $^1$H and $^{207}$Pb, respectively. While the
  $^{207}$Pb-NMR signal was wiped out at low temperatures due to very short spin-spin and spin-lattice relaxation
  rates, the $^1$H-NMR signal could be observed in the whole temperature
  regime. Note that the average 1/$T_1$-value has been extracted for $H$ $||$ $b$ due to overlapping $^1$H-NMR lines. The lines are guides to the eyes.}
  \label{spinlattice}
\end{figure}

The temperature dependence of the spin-lattice relaxation rate,
1/$T_1$, is depicted in Fig.~\ref{spinlattice} for both probing
nuclei and for the external magnetic field parallel to the $a$, $b$,
and $c$ axes. Here, we present our spin-lattice relaxation data in
both the paramagnetic and the magnetically ordered state below
$T_\mathrm{N}$. This way, a first microscopic proof for the 3D
magnetically ordered state is given. Due to the very short spin-spin
and also spin-lattice relaxation times of the order of 5 and
20\,$\mu$s, respectively, in the temperature range close to
$T_\mathrm{N}$, our data are marked by large error bars in this
temperature region and an accurate determination of $T_\mathrm{N}$
via 1/$T_1$ is difficult. From the overall temperature dependence of
the spin-lattice relaxation rate $T_\mathrm{N}$ was determined to be
about (2.5$\pm$0.2)\,K at 2\,T.

While the longitudinal nuclear magnetization is well described by
the standard expression of a nuclear spin $I$ = 1/2 with a single
$T_1$ component for $^{207}$Pb and for all field directions, for
$^1$H an additional $T_1$ component with a very short spin-lattice
relaxation time of the order of $\sim$ 10 $\mu$s needs to be taken
into account for the whole temperature range for $H$ $||$
$b$.~\cite{note3} Although it cannot be ruled out completely that
the short $T_1$-component arises from either impurities in the
sample or from a $^1$H background signal from e.g. teflon outside
the coil, both its qualitatively similar temperature dependence as
well as its complete absence for the other two directions, strongly
hint towards an intrinsic signal. However, the origin of this very
fast-relaxing component for $H$ $||$ $b$ is not clear up to now and
needs further investigation. Thus, we solely concentrate on the
single, longer $T_1$ component in the following.

Overall, as temperature is lowered, for both nuclei one finds a
strong increase in 1/$T_1(T)$ with a sharp peak at $T_\mathrm{N}
\approx$ 2.5(2)\,K, below which the spin-lattice relaxation rate
decreases again as shown for all three principle directions
(Fig.~\ref{spinlattice}). This sharp peak is indicative of a
transition into the 3D ordered state. The divergence of the
relaxation rate 1/$T_1$ due to the transition at $\sim$ 2.5\,K makes
a further comparison to any low-dimensional (1D, 2D) models in the
paramagnetic regime difficult, as it masks the quasi-1D behavior of
linarite at low temperatures even above $T_\mathrm{N}$. However, it
is crucial to emphasize that the $^1$H-NMR investigations in the 3D
ordered state have shown a sharp splitting into a multi-peak
pattern, which is consistent with a non-trivial antiferromagnetic
alignment of the Cu-spins below $T_\mathrm{N}$. Correspondingly, the
1/$T_1$ data below $T_\mathrm{N}$, shown in Fig.~\ref{spinlattice}
as closed circles, reflect the average value of 1/$T_1$ for all
lines with the small distribution of values marked by error bars.

\section{Theoretical Analysis - assignment of exchange integrals}

\subsection{Saturation fields and Curie-Weiss temperature}

As a starting point, we assume at first that the interchain
interaction is weak and that the inchain interaction (frustration)
obeys the rigorous two-magnon bound state condition \cite{kuzian07}
$\alpha > \alpha_{\rm 3c}= 0.3676776$, i.e., it is within the
quadrupolar region at the saturation field. Then, using the
experimental value for the Curie-Weiss temperature
$\Theta_{\mbox{\tiny CW}} \approx 0.5|J_1|(1-\alpha) \approx 27$~K,
reported above, the estimated 1D saturation field of about
$H^a_s\approx 5.5$~T ($H_{\rm c3}$, see Fig.\ 3), and the $g$-factor
for $H \parallel a$, i.e., $g_a=2.34$ (see Sect.\ IV), we obtain for
the ratio $r$ of the former quantities for the $a$ direction

\begin{equation}
r=\frac{\Theta_{\mbox{\tiny CW}}} {g\mu_{\mbox{\tiny B}} H_s}=
\frac{1-\alpha^2}{4\alpha^2+2\alpha -1} \approx 3.122575.
\label{rfac}
\end{equation}
\noindent Note, that the last equation is a rigorous relation valid
in the commensurate \cite{remq,kuzian07} 1D quadrupolar phase (QP)
of the adopted spin model. Inverting Eq.\ (\ref{rfac}), we obtain
for the frustration ratio

\begin{equation}
\alpha (r) =\frac{\sqrt{5+5/r+1/r^2}-1}{4+1/r}\approx 0.36784
\label{a4}
\end{equation}
\noindent still in the quadrupolar phase, although very close to the
border to the octupolar phase at $\alpha_{\rm 3c}$. From
$\Theta_{\mbox{\tiny CW}}$ we obtain $J_1=-85.4$~K and $J_2=31.4$~K.

In the opposite limit of strong enough antiferromagnetic interchain
coupling, where multipolar effects disappear, \cite{nishimoto11} the
former can be described at least approximately by the well-known
one-magnon theory in the case of separable in-chain and interchain
contributions to the total 3D (2D) saturation field. \cite{kuzian07}
As a result one arrives at

\begin{eqnarray}
r&=&
0.5\frac{1-\alpha -\beta }{2\alpha -1+0.125/\alpha}, \\
\alpha &
= &
\frac{0.5}{4r+1}\left[2r+1-0.5\rho
+\sqrt{(3-2\rho )r
+(1-0.5\rho )^2}\right] \nonumber \\
& \approx &0.3639,  \\
\nonumber
\end{eqnarray}
where $\beta$ denotes the 2D interchain coupling in the basal plane
measured in units of $|J_1|$ and $\rho=H^{\rm 3D}_s/H^{\rm
1D}_s-1\approx 0.3819$ is given by the ratio of the 3D(2D) and the
1D saturation fields, 7.6~T and 5.5~T, respectively, and the
experimental value $r=3.122575$ has been used. These numbers result
in $J_1=-86.7$~K and $J_2=31.6$~K. Notice that the obtained value
$\alpha$ is very close to the "two-magnon-value" estimated above and
the value obtained from the analysis of the magnetization
measurements (see Fig.\ 14).

\subsection{Spin susceptibility and magnetization curve}

Noteworthy, from the maximum position of the spin susceptibility at
$T^{\rm max}_\chi=4.9\pm 0.3$~K in units of $J_2$ or $|J_1|$ (Fig.\
13) we can also estimate an $\alpha$-value for our compound. Here,
$T^{\rm max}_\chi$ is described with high precision by

\begin{equation}
\frac{T^{\rm max}_{\chi}(\alpha) }{|J1|}= \sum ^6_{m=1}
A_m\left(\alpha -\frac{1}{4}\right)^m \quad \mbox{for} \quad 9/4
\geq \alpha \geq 1/4 , \label{Tmaxchi}
\end{equation}

\begin{figure}[t]

\begin{center}
\includegraphics[width=0.75\columnwidth]{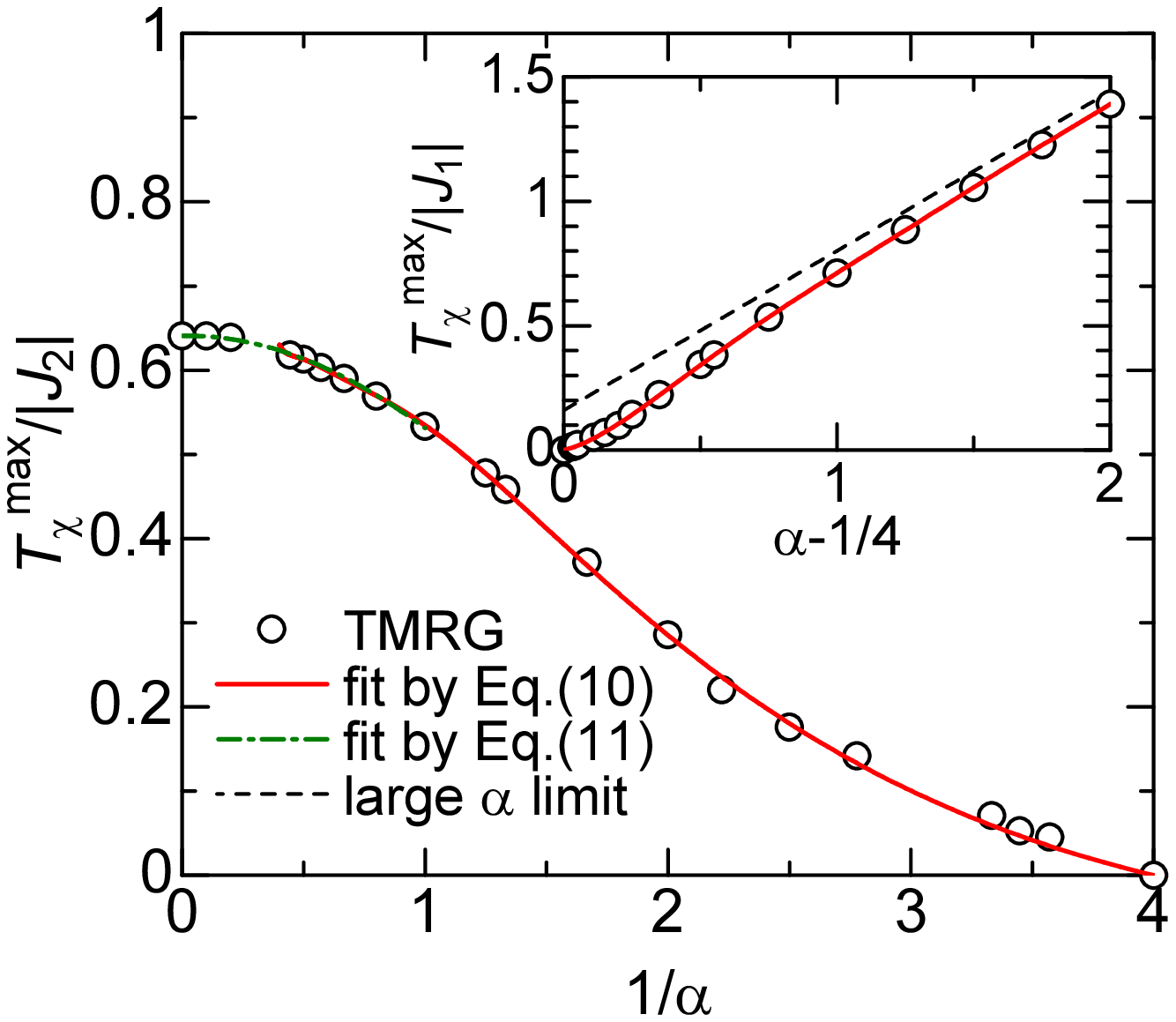}
\end{center}
\caption[2]{(Color online) Maximum position $T^{\rm max}_{\chi }$ of
the spin susceptibility $\chi(T)$ measured in units of $J_2$ as a
function of $1/\alpha$, i.e., viewing the $J_1$-$J_2$ model as an
equivalent realization of two interpenetrating {\it interacting} AFM
Heisenberg chains. In the inset the maximum position $T^{\rm
max}_{\chi}$ of $\chi(T)$ measured in units of $|J_1|$ vs.\ the
frustration parameter $\alpha$ according to DMRG calculations is
shown (see text). The broken line describes the asymptotic curve
$T^{\rm max}_{\chi } =0.641 J_2 \equiv 0.641\alpha |J_1|$, well
known from the Bethe-Ansatz-based solution for the unfrustrated AFM
spin-1/2 Heisenberg chain. \cite{johnston97}} \label{peakpos}
\end{figure}

\noindent with $A_1=0.2778$, $A_2=1.7055$, $A_3=-2.559$,
$A_4=1.8487$, $A_5=-0.6499$, and $A_6=0.0891$.

The expression given by Eq.\ (\ref{Tmaxchi}) has been obtained from
a fit of our TMRG-data for $\chi(T)$ for strong and intermediate
coupling. We note that Eq.\ (10) also describes the experimental
situation and parametrizations suggested for Li$_2$ZrCuO$_4$ with
$\alpha$ = 0.3~\cite{drechsler07} as well as for LiVCuO$_4$ with
$\alpha$ = 0.75.~\cite{drechsler11,nishimoto11} In the opposite
weak-coupling limit it is convenient to expand $T^{\rm max}_{\chi}$
around the limiting point of decoupled interpenetrating AFM
Heisenberg chains in powers of $1/\alpha=|J_1|/J_2$:

\begin{equation}
\frac{T^{max}_\chi (\alpha)}{J_2}=0.641-\sum_{m=2}^5
\frac{D_m}{\alpha^m}, \qquad \mbox{for} \quad \alpha \geq 1,
\end{equation}
\noindent with $D_2$ = 0.0034, $D_3$ = 0.499, $D_4$ = $-$0.669, and
$D_5$ = 0.281.

For linarite, Eq.\ (10) yields somewhat larger values for the same
$\alpha$, namely $J_1$ = -107.5\,K and $J_2$ = 39.5\,K. We ascribe
this difference to the presence of an antiferromagnetic interchain
coupling ignored in the former estimation and an uncertainty in the
determination of $\Theta_{\mbox{\tiny CW}}$. (Note that $T^{\rm
max}_{\chi }$ is not affected by the interchain coupling within the
random phase approximation adopted for its treatment at variance to
the Curie-Weiss temperature $\Theta_{\mbox{\tiny CW}}$.)

A completely different situation has been suggested by Y.\ Yasui
{\it et al.} [\onlinecite{yasui11}]. These authors arrived at a
$T^{\rm max}_\chi \approx 3.8$ to 5\,K somewhat lower than our
values but at a {\it negative} $\Theta_{\mbox{\tiny CW}} \approx
-4$~K, $J_1=(-13 \pm 3)$~K, and an NNN exchange integral $J_2=(21\pm
5)$~K, yielding a frustration ratio $\alpha$ = 1.6 belonging to the
weak-coupling region. Adopting their $\alpha$-value and using their
experimental $T^{\rm max}_\chi$ of about 5~K (see inset in Fig.\ 2
of Ref.\ \onlinecite{yasui11}), one would arrive at $J_2 \approx
8$~K and at $J_1 \approx -5.2$~K, which is inconsistent with (21$\pm
5$)~K and ($-13\pm 3$)~K, respectively, claimed before.

\begin{figure}[b]
\begin{center}
\includegraphics[width=0.8\columnwidth ]{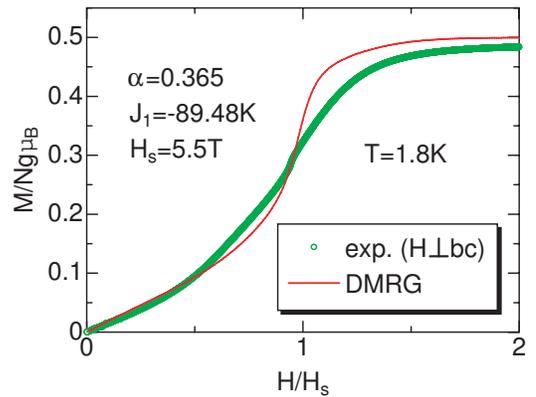}
\end{center}
\caption{(Color online) Fit for the magnetization vs.\ external
field of PbCuSO$_4$(OH)$_2$ at $T$ = 1.8\,K and for $H$ $||$
$a_\perp$.} \label{fig:mag1D}
\end{figure}

Next, we checked these preliminary values of the in-chain exchange
integrals by additionally describing the low- and intermediate-field
magnetization data at $T$ = 1.8\,K for the $a_\perp$ direction
(Fig.~\ref{fig:mag1D}). The high-field region has not been taken
into account due to an additional phase transition of yet unknown
nature at low temperatures. The best fit at low fields is obtained
for $\alpha$ $\sim$ 0.365. Then, the corresponding $J_1$-value
consistent with a 1D saturation field of $H_s$ = 5.5\,T becomes
$J_1$ = -89.5\,K. These value are in accord with the values
estimated above. Therefore they might be regarded as first realistic
phenomenological values despite the ignored but certainly present
weak interchain and spin-anisotropy effects, only briefly discussed
below as an outlook for future studies.

\begin{figure}[t]
\includegraphics[width=0.84\columnwidth, angle=-90]{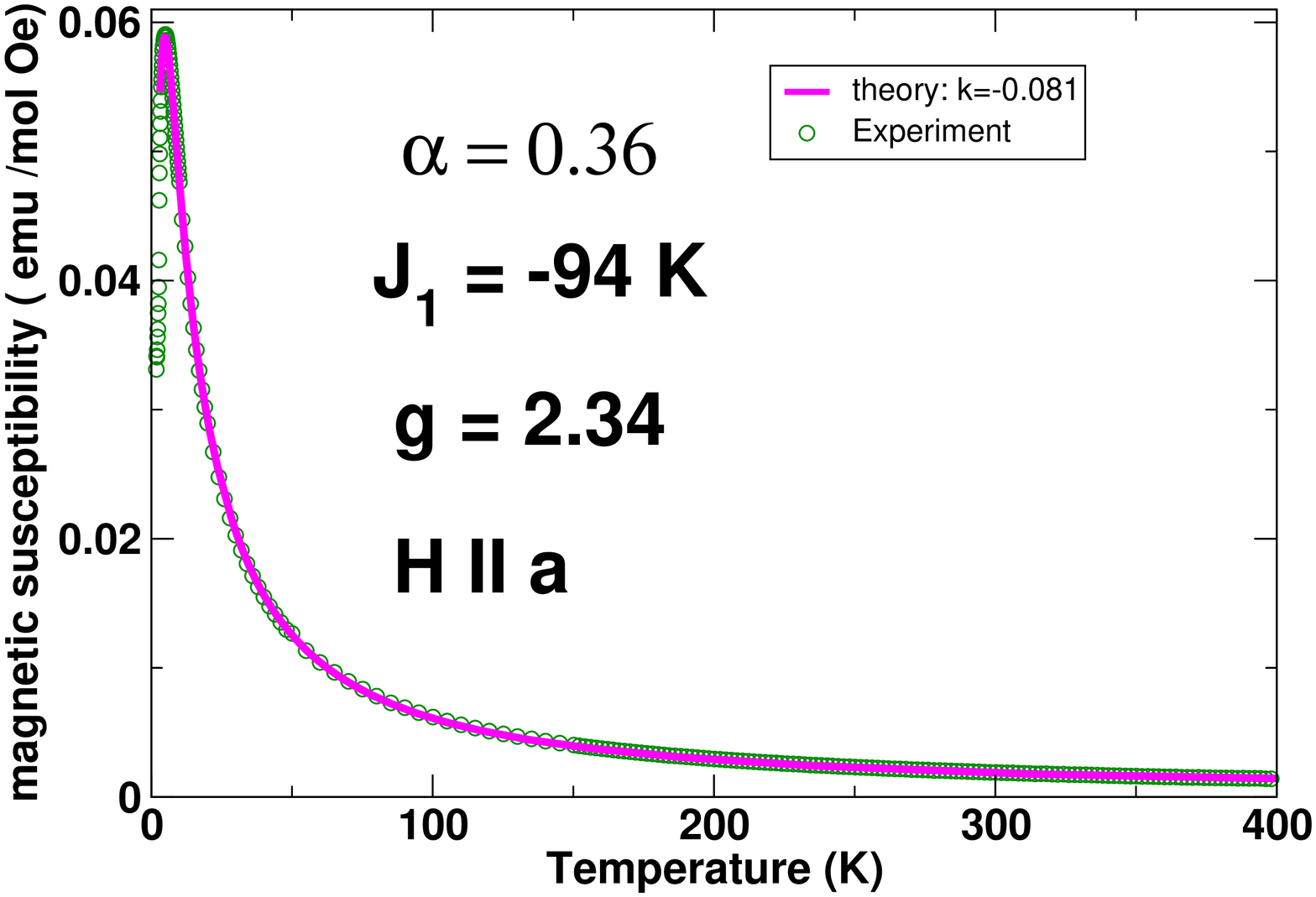}
\includegraphics[width=0.84\columnwidth, angle=-90]{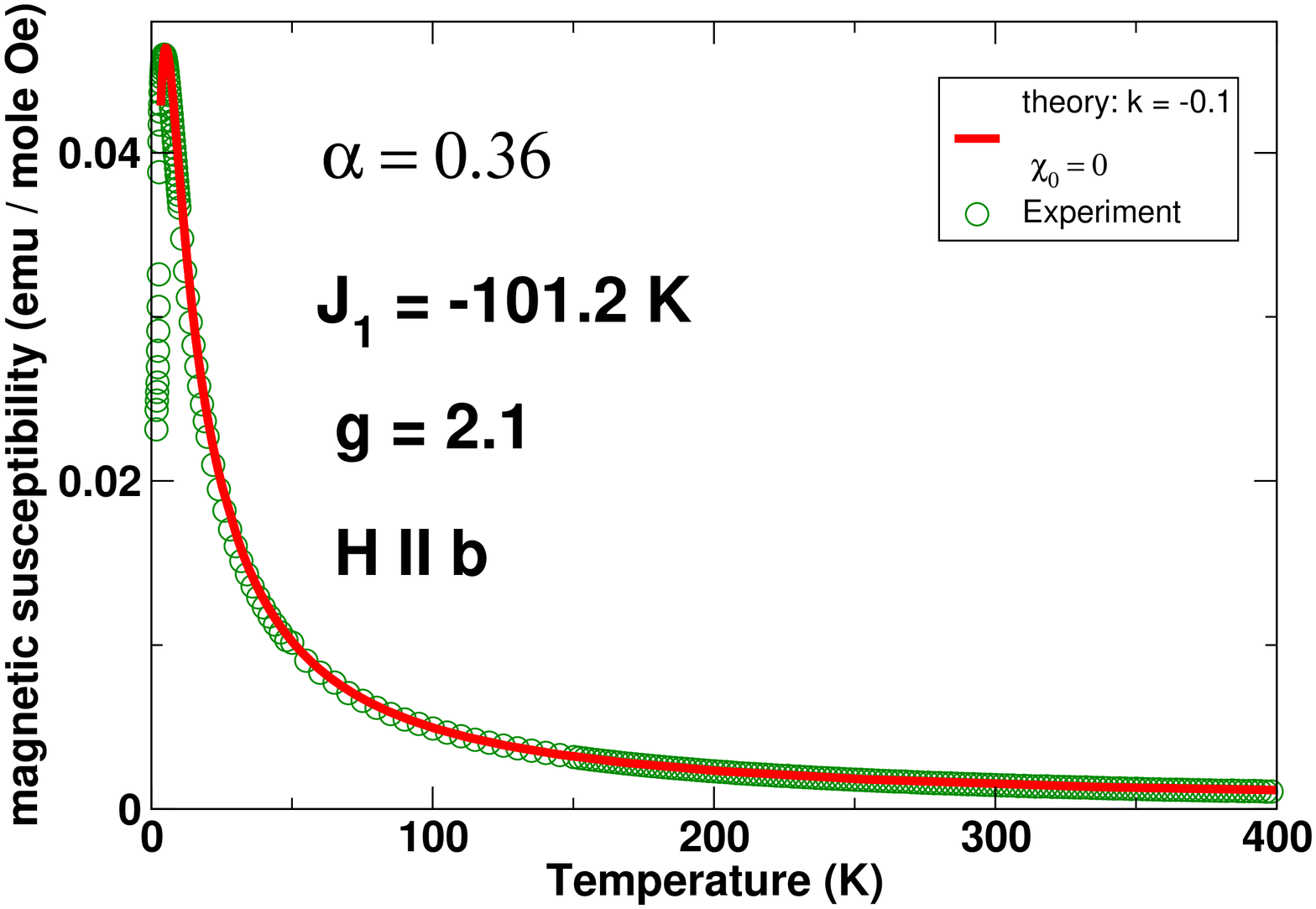}
\caption{(Color online) Analysis of  the magnetic susceptibility,
$\chi(T)$, within the isotropic $J_1$-$J_2$ model. The fitted solid
lines result in slightly different IC parameters $k= -0.081$ to -0.1
(see text) corresponding to a weak FM IC of a few K. (a) $H
\parallel a$, (b) $H \parallel b$ axis.
The $g$-factors are taken from Sect.\ IV.
} \label{chic}
\end{figure}

Finally, the analysis of the spin-susceptibility data nicely
confirms the above results (Fig.\ \ref{chic}). In the analysis the
obtained experimental parameters for the $g$-factor and the
Curie-Weiss constant have been used. The best matching between
experimental data and theoretical calculations results in magnetic
exchange parameters $J_1$ = -94\,K and $J_1$ = -101.2\,K ($\alpha$ =
0.36) for the $a$ and $b$ direction, respectively, again
highlighting the magnetic anisotropy in our system.

In summary, within the simplest isotropic spin chain model our
values for $J_1\approx -97\pm 10$~K and $J_2\approx 36\pm 4$~K
exceed significantly those determined previously from fits of the
$T$-dependent susceptibility in the high-$T$ regime, where only
$J_1$ = $-$30~K (-13~K) and $J_2$ = 15~K (21~K), respectively, have
been extracted.~\cite{baran06,yasui11} In addition, we also found a
considerably smaller frustration parameter $\alpha \approx 0.37$
instead of 0.5 and 1.62, respectively,\cite{baran06,yasui11} which
puts linarite closer to the 1D critical point at 0.25 with
consequences for a weaker critical antiferromagnetic interchain
coupling for ordered multipolar phases at $T=0$.\cite{nishimoto10}
The former might be masked by non-negligible impurity contributions
in accord with a 3D analysis of the anisotropic susceptibility data
(see below).

\subsection{Aspects of interchain coupling and symmetric exchange
anisotropy for $\chi(T)$}

The interchain coupling (IC) has been taken into account in the
frame of the frequently used random phase approximation (RPA):
\cite{johnston97,bocquet01}

\begin{equation}
\chi_{\mbox{\tiny 3D}}(T)\approx \frac{\chi_{\mbox{\tiny 1D}}}
{1+k\chi_{\mbox{\tiny 1D}}(T)},
\end{equation}
where the one-dimensional susceptibility, $\chi_{\mbox{\tiny
1D}}(T)$, has been calculated applying the TMRG-method, the
interchain coupling $k=(g/2)^2\sum J_{ic}/|J_1|$ yields a
temperature independent single parameter and the summation runs over
all interchain couplings. The fits to the data shown in Fig. \ 15
result in slightly different IC parameters $k= -0.081$ and -0.1 for
the $a$ and $b$ direction, respectively, corresponding to a weak FM
IC of a few K. In principle, a small underestimation of $\chi(T)$
within its maximum region can be removed adopting also an impurity
contribution described by a Curie or a Curie-Weiss law as frequently
used in the literature for other quasi-1D compounds (see e.g.\
Refs.\ \onlinecite{johnston97,nath05}).
Despite the uncontrolled deviations introduced by the RPA, this way,
the determination of the interchain coupling is not unique and a
(sizable) impurity contribution might mask weak antiferromagnetic
interchain interactions.

A weak AFM IC of the order of $J_{ic} \approx $ 4~K follows also
from the positive difference between the estimated 3D saturation
fields and its 1D counter part of $\approx$ 6~T.~\cite{kuzian07}
Alternatively, the high-field magnetization should be affected
significantly by changed exchange integrals in the new high-field
phase observed above $\sim$ 6~T. Inelastic neutron data such as in
Ref.\ [\onlinecite{lorenz10}] might be useful to achieve a more
precise assignment of the details such as the principle nature of
the interchain coupling in our compound.

Due to the increase of a possible, small impurity contributions with
decreasing $T$ in our compound, the maximum positions of the
impurity corrected susceptibilities are upshifted by a few tenths of
a K. Naturally the effect is largest for the $b$-axis susceptibility
(shift $\sim$ 0.5\,K). Nevertheless, a significant anisotropy
remains also for these "corrected" susceptibilities which points
together with the obtained largest exchange integral $J_1$ for the
$b$ axis to a corresponding easy-axis assignment. To improve further
the theoretical analysis, a detailed consideration of an anisotropic
$J_1$-$J_2$ model would be desirable which, however, is outside the
scope of the present paper. For a first estimate see Fig.\ 16 and
the short discussion given below.

\begin{figure}[t!]
\begin{center}
\includegraphics[width=0.84\columnwidth,angle=-90]{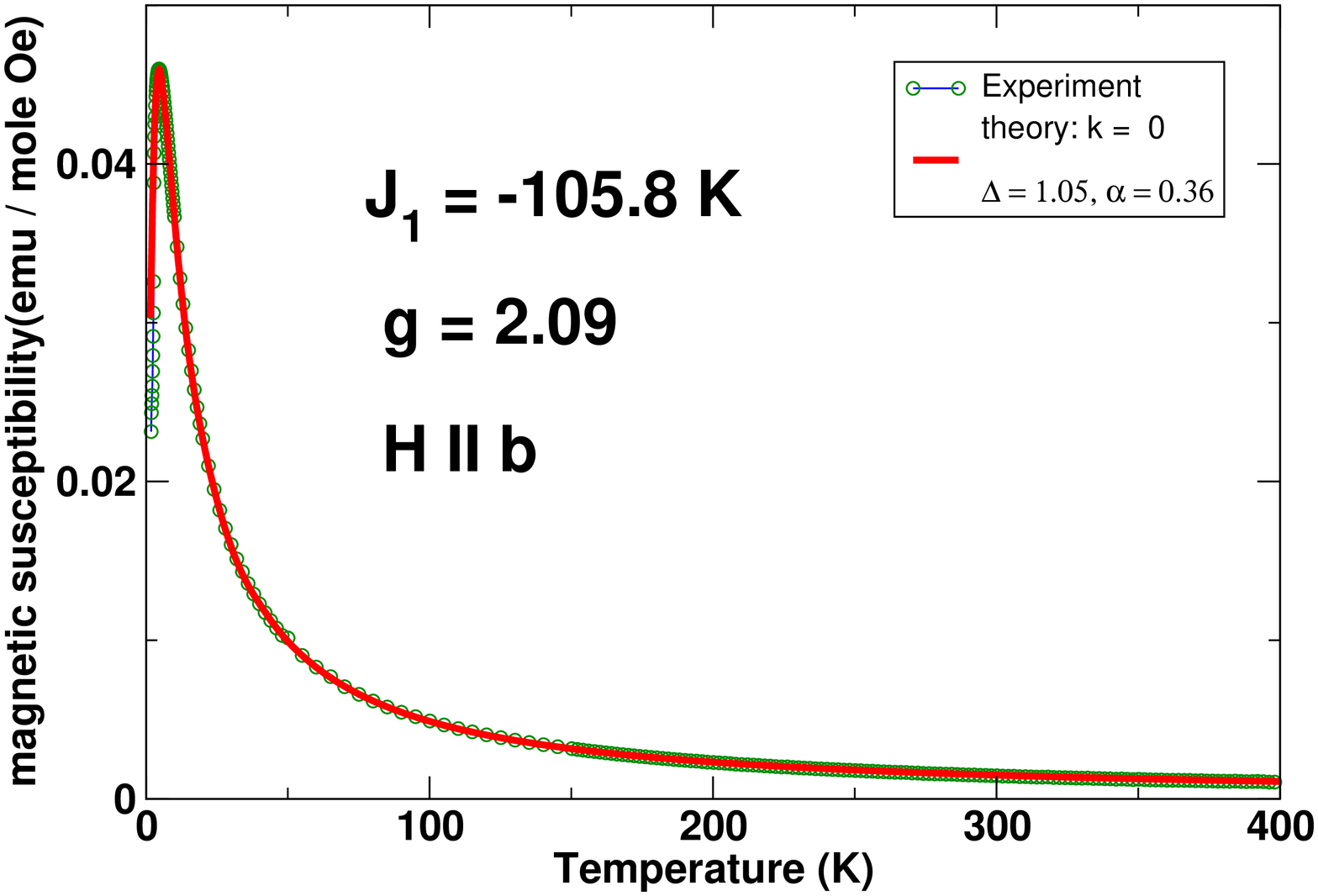}
\includegraphics[width=0.84\columnwidth,angle=-90]{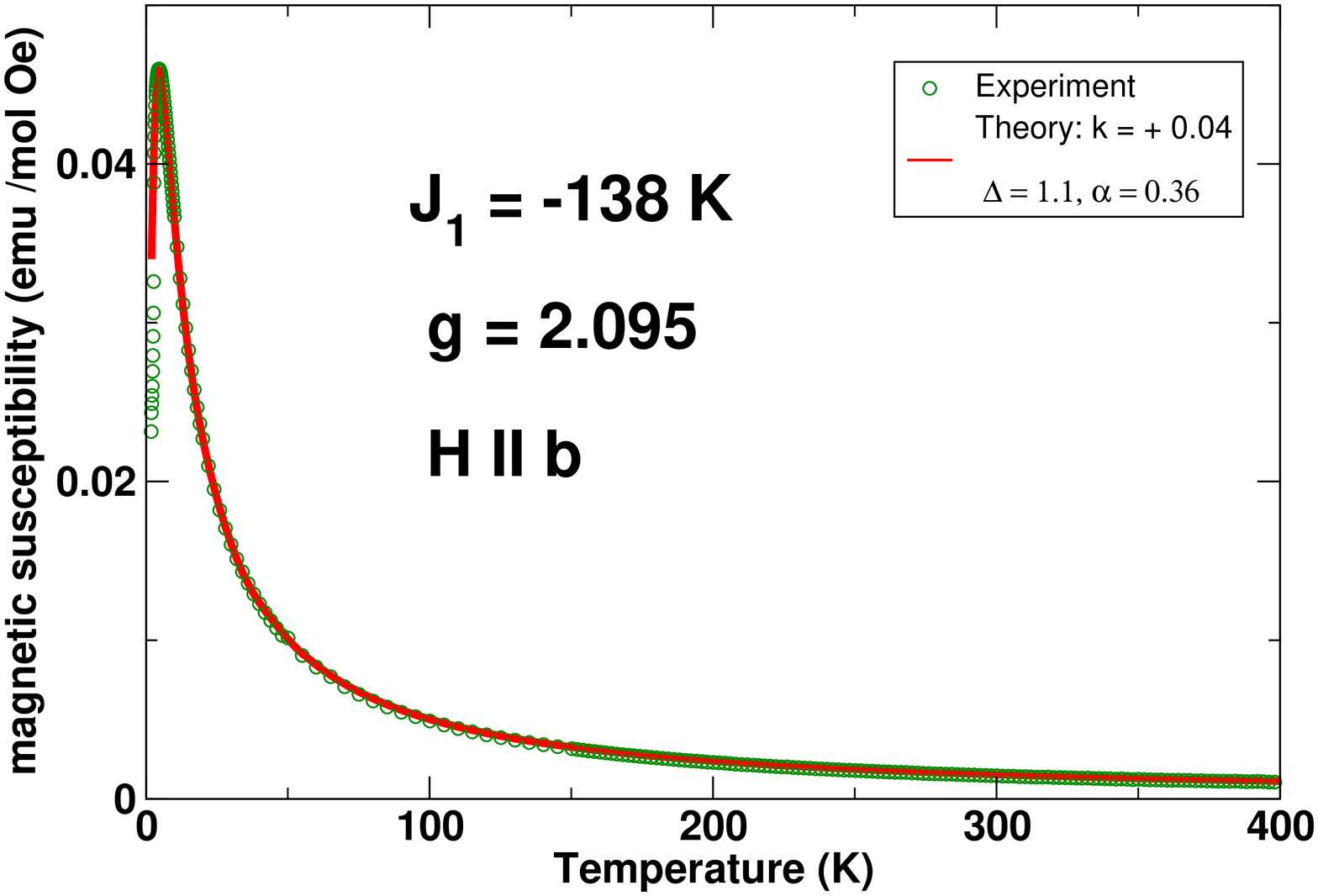}
\end{center}
\caption{(Color online) Magnetic susceptibility for a magnetic field
applied along the suggested easy-axis ($b$ axis) fitted within the
1D anisotropic $J_1$-$J_2$ model based on TMRG calculations
supplemented by zero (upper panel) and finite antiferromagnetic
(lower panel) interchain interactions treated in the RPA. The
adopted easy-axis anisotropy is measured by the dimensionless
parameter $\Delta > 1$ (see text).} \label{eaxis}
\end{figure}

\begin{figure}[b!]
\begin{center}
\includegraphics[width=0.84\columnwidth]{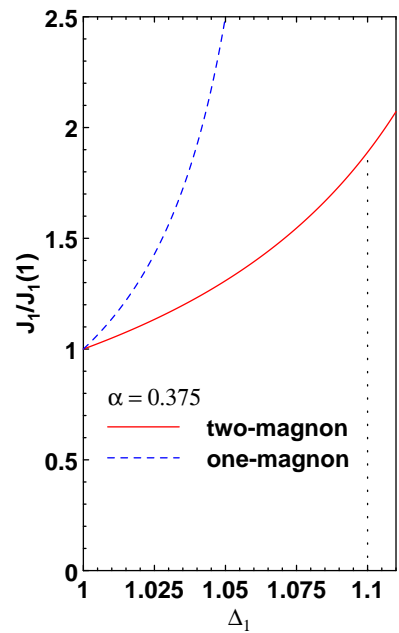}
\end{center}
\caption{(Color online) Value of the isotropic part of the
ferromagnetic nearest-neighbor exchange integral $J_1$ in units of
the corresponding value in the isotropic model vs. the dimensionless
anisotropy factor $\Delta_1 \geq 1$ as derived from Eqs.\
(\ref{hsataniso}) and (\ref{onemaganiso}).} \label{J1aniso}
\end{figure}

Noteworthy, the quality of the fits can be improved without adopting
a Curie-Weiss like impurity contribution, if instead a model with
symmetric exchange anisotropy is applied. According to previous
theoretical investigations of edge-shared cuprate chains with a
ferromagnetic nearest neighbor exchange coupling, an easy-axis
exchange anisotropy is expected.~\cite{aharony,hayn} Therefore the
first term in the isotropic spin-Hamiltonian should be replaced by

$$J_1\vec{S}_i\vec{S}_{i+1} \rightarrow
J_1\left(
S_i^xS_{i+1}^x+S_i^zS_{i+1}^z
\right)
+J_1\Delta_1S
^{y}_i
S^y_{i+1},
$$
with $\Delta_1 > 1$ (see Fig.\ \ref{eaxis}). Such an easy-axis
exchange anisotropy affects significantly the low-temperature
behavior of $\chi(T)$ and the saturation field in the low
$\alpha$-region of interest. Following Ref.\ \onlinecite{kuzian07},
but ignoring weak anisotropy effects for the next-nearest neighbor
exchange (i.e., setting $\Delta_2=1$), we rewrite the leading term
in the expression for the 1D saturation field valid in the
two-magnon case at weak interchain coupling as

\begin{equation}
g\mu_{\rm B}
H_s(\Delta_1)=|J_1|\left(
2\alpha -\Delta_1+0.5\Delta^2_1/\left[\Delta_1+\alpha\right]
\right),
\label{hsataniso}
\end{equation}
and analogously in the one-magnon case:
\begin{equation}
g\mu_{\rm B}
H_s(\Delta_1)=|J_1|\left(
2\alpha -\Delta_1
+
0.125/\alpha
\right).
\label{onemaganiso}
\end{equation}
The effect of the easy-axis anisotropy under consideration on the
in-chain exchange coupling $J_1$ is illustrated in
Fig.~\ref{J1aniso} for the case we are interested in here, namely,
when one extracts the $J_1$-value from a given experimental
saturation field $H_s$.

In the presence of an antiferromagnetic interchain coupling the
enhancement effect for the corresponding renormalization of $J_1$ is
a bit less dramatic but nevertheless significant. Such a behavior is
in accord with a similar effect found for the magnetic
susceptibility data fitted by the isotropic and the aniosotropic
model as shown in Figs.~\ref{chic},\ref{eaxis}. Thus, within such a
scenario the observed increase of $|J_1|$ by more than 30~K becomes
rather natural. These results are in reasonable agreement with
recent L(S)DA+$U$ calculations, based on a refined crystallographic
structure of PbCuSO$_4$(OH)$_2$ including the corrected proton
positions,~\cite{schofield09} see below. Note that the range of
validity of the expression given by Eq.\ (\ref{hsataniso}) with
respect to the formation of three-magnon bound states is under
investigation at present. However, qualitatively the same behaviour
is also expected for the octupolar three-magnon and other multipolar
phases. A systematic study of exchange anisotropy effects including
also $J_2$ as well as the interchain coupling is postponed for a
future study.

\subsection{Density functional theory: L(S)DA + $U$}
\begin{table}[b!]
\begin{tabular}{|l|c|c|}
\hline axis & $a$ & $b$ \\ \hline
\hline $g$ (ESR) & 2.34 & 2.1  \\
\hline $\Theta_{CW}$ (K) & 27  & 27  \\
\hline $H_{\rm s}$(T) from & 5.5 & \\
1D $M(H,T=1.8\mbox{K})$ &&\\
\hline $-J_1$ (K) (IHM, $\chi(T)$) & 94 & 101  \\
\hline $J_2$ (K), (IHM, $\chi(T)$) & 33.8 & 36.4 \\
\hline $-J_1$ (K) (IHM, $M(H)$)& 89.5&   \\
\hline $J_2$ (K) (IHM, $M(H)$)& 32.7&   \\
\hline $-J_1$ (K), (AHM) &  & 138  \\
\hline $\alpha$ (IHM, $M(H)$)& 0.365 &  \\
\hline $\alpha$ (AHM) &  & 0.36  \\
\hline -$J_{1}$ (K), (LSDA+U) & 133 & 133 \\
\hline $J_{2}$ (K), (LSDA+U) & 42 & 42 \\
\hline
\end{tabular}
\caption{Calculated magnetic exchange interactions $J_1$ and $J_2$
along the chain using a 1D approach based on the analysis of
saturation fields at $T$ = 0, low- and intermediate field
magnetization data within the DMRG, and LSDA + U calculations,
together with the experimental values for the $g$-factor, $H_s$, and
$\Theta_{CW}$ of PbCuSO$_4$(OH)$_2$; for details see text. The
abbreviations IHM and AHM denote isotropic and anisotropic
Heisenberg model, respectively. }~\label{tab:alpha}
\end{table}

\begin{figure}[b]
\begin{center}
\includegraphics[width=0.75\columnwidth]{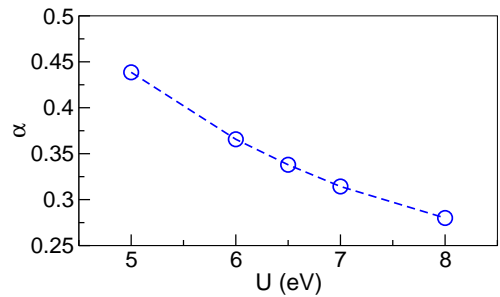}
\end{center}
\caption{Frustration parameter vs.\ $U$ from L(S)DA+$U$ for
PbCuSO$_4$(OH)$_2$. The Hund's rule exchange at Cu sites has been
fixed at $J$ = 1.0~eV for all calculations.} \label{U}
\end{figure}

To probe our new parameter set with respect to a microscopic
picture, we carried out DFT band structure calculations within the
LSDA+$U$ scheme which takes into account the strong Coulomb
repulsion, $U$, at the Cu site. Since we observed a sizable
dependence of the resulting exchange parameters from the H
position,~\cite{schmitt11} our calculations are based on the
recently refined crystal structure of Ref.~\onlinecite{schofield09}.
For a screened Coulomb repulsion $U$ = 7 eV, which is in the typical
range of $U$ values that were successfully applied to related Cu-O
systems,\cite{johannes06,lorenz10,schmitt09} and the usual value of
the Hund's rule coupling $J=1.0$~eV which enters the L(S)DA+$U$
calculational scheme, we obtain a frustration ratio $\alpha \approx$
0.32. This value is in reasonable agreement with the value $\alpha
\approx 0.36$ derived from the experimental data (compare Tab.\ 1),
in view of possible renormalizations due to a non-negligible
spin-lattice coupling in the non-adiabatic limit or intermediate
case and of possible quantum-effect caused by the zero-point motion
of the light hydrogen nuclei ignored in all density functional
approaches.

The resulting frustration ratio $\alpha$ depends only moderately on
$U$ within the physically reasonable range between 5 and 8 eV (Fig.\
\ref{U}). The respective calculated exchange integrals ($U$ = 7 eV)
are $J_1$ = -133 K and $J_2$ = 42 K. These numbers are in good
agreement with the isotropic parts obtained within the easy-axis
model for the fit of the magnetic susceptibility with $H \parallel $
$b$ reported above: $-J_1=138$~K and $J_2=49.7$K. In other words,
also the LSDA+$U$ derived exchange integrals clearly support a
scenario with significantly larger $J$- values compared to those
proposed in Refs.~\onlinecite{yasui11,baran06}. Taking into account
the different interchain couplings $J_{ic,0}$, $J_{ic,1}$, and
$J_{ic,2}$ (see Fig. 1) as well as the respective number of
neighbors we can estimate an effective interchain coupling $J_{ic}
\sim$ 7 K. This is also consistent with the analysis of $\chi(T)$
within the easy-axis model which yields $\approx$ 5~K. A detailed
electronic-structure study of the title compound under
consideration, in particular its structural stability and the
sizable dependence of the coupling parameters from the details of
the crystal structure, will be published elsewhere.\cite{schmitt11}

\section{Conclusion}

In conclusion, we have performed a detailed combined experimental
and theoretical study of PbCuSO$_4$(OH)$_2$ in the paramagnetic
regime in fields up to saturation. The saturation field, probed via
magnetization studies at low temperatures, is anisotropic for the
three principal crystallographic axes ranging between 7.6 and
10.5\,T. ESR and NMR measurements further prove that the static
susceptibility is dominated by the intrinsic spin susceptibility.
The Knight shift as well as the broadening of the linewidth in the
ESR and NMR data at elevated temperatures indicate a frustrated
system with the onset of anisotropic magnetic correlations far above
the magnetic ordering temperature, $T_\mathrm{N}$ = 2.75(5)\,K. Our
experimental data are analyzed both in 1D as well as quasi-1D
approaches based on DMRG and hard-core boson calculations, yielding
values for the exchange interactions $J_1 \sim -100$~K and $J_2 \sim
36$~K along the chain as well as a weak interchain coupling of a few
K, leaving room for a quadrupolar phase ({\it i.e.}, spin nematics)
in experimentally accessible magnetic fields.

In view of the small absolute value of the saturation field and the
rich manifold of field-induced phases (particularly along the chain
direction), PbCuSO$_4$(OH)$_2$ reveals itself as a promising
material for additional investigations in the future. In particular,
linarite possibly realizes a prototype for investigating the
recently predicted spin multipolar order close to the saturation
field, $H_s$, of such frustrated "ferromagnetic" spin-chain
compounds.

From a theoretical point of view, such a systematic study of one
compound on high-quality single crystals might be very helpful for a
future unified theory of frustrated edge-shared cuprates in order to
understand what is really puzzling in each compound of that
increasing family and what some of these compounds have in common.
In particular, the question which of their properties can be used
best for the determination of a specific model parameter from
experimental data seems to be important.

\section*{Acknowledgements}

We acknowledge fruitful discussions with R.\ Giraud, H.-J.\ Grafe,
C.\ Berthier, R.\ Kuzian, J.\ Richter and R.\ Zinke. We thank O.\
Kataeva, S.\ Partzsch and J. Geck for the x-ray characterization of
the samples and S.~Ga{\ss} for technical support. We are
particularly thankful to G. Heide and M. G\"{a}belein from the {\it
Geoscientific Collections} for supplying us with their linarite
crystal of \textit{origin 1}. F. Pfeiffer is acknowledged for
supplying us with the linarite samples of \textit{origin 2}. This
work has been supported by the DFG through the grants: WO 1532/3-1,
SU 229/10-1, FOR 912 and DR 269/3-1. Part of this work has been
supported by EuroMagNET under the EU contract no. 228043.


\begin{references}
\bibitem{note1} Due to technical reasons (sample mounting and sample space)
the field was applied perpendicular to the $(bc)$ plane rather than
parallel to the $a$ axis for the magnetization measurements.
However, since the angular dependence of $M$ is expected to be
proportional to $\cos\beta$ ($\beta$ = angle between $H$ and the
principal crystallographic direction), a misalignment of about
12$^\circ$ with respect to the $a$ axis only results in a change of
$M$ by less than 3$\%$.
\bibitem{noteESR} Note that resonance fields are as measured and not
corrected by the temperature-dependent resonance frequency of the
cavity, which can cause a shift of the order of 10\,mT from 100\,K
to 300\,K.
\bibitem{note2} The practically unavoidable simultaneous excitation of
the two inequivalent $^1$H-NMR lines at high temperatures prevented
an exact determination of the NMR shift in the high-temperature
regime above 200-250\,K. Therefore, only $^{207}$Pb-NMR was
conducted in this temperature regime.
\bibitem{note3} Since such a short spin-lattice relaxation time is far
too small to be determined quantitatively considering the deadtime
of our spectrometer, we used a saturation recovery sequence with an
echo subsequence at the end for the determination of the
spin-lattice relaxation time for $H$ $||$ $b$. In this way, the
short $T_1$ component was revealed by a small non-zero magnetization
component for $\tau_{var}$$\rightarrow$ 0 and could be neglected in
the following analysis of the second, larger $T_1$ component.
\end{references}
\end{document}